\documentclass[%
 aip,
 jmp,%
 amsmath,amssymb,
 reprint,%
]{revtex4-2}

\usepackage{graphicx}
\usepackage{dcolumn}
\usepackage{bm}
\usepackage{verbatim} 

\usepackage{color}
\definecolor{orange}{rgb}{0.95,0.5,0.05}
\definecolor{shiningblue}{rgb}{0.3,0.68,0.89}
\definecolor{othergreen}{rgb}{0.05,0.7,0.02}

\begin{document}

\title{Traveling fronts in self-replicating persistent random walks with multiple internal states}

\author{Keisuke Ishihara}
\email[]{kishihara@pks.mpg.de}
\affiliation{Max Planck Institute for the Physics of Complex Systems, N\"{o}thnitzer Strasse 38, 01187 Dresden, Germany}
\affiliation{Max Planck Institute of Molecular Cell Biology and Genetics, Pfotenhauer Strasse 108,
01307 Dresden, Germany}

\author{Ashish B. George}
\affiliation{Department of Physics, Boston University, Boston, MA 02215}

\author{Ryan Cornelius}
\affiliation{Department of Physics, Boston University, Boston, MA 02215}

\author{Kirill S. Korolev}
\email[]{korolev@bu.edu}
\affiliation{Department of Physics, Boston University, Boston, MA 02215}
\affiliation{Graduate Program in Bioinformatics, Boston University, Boston, MA 02215}

\date{\today}

\begin{abstract}
Self-activation coupled to a transport mechanism results in traveling waves that describe polymerization reactions, forest fires, tumor growth, and even the spread of epidemics. Diffusion is a simple and commonly used model of particle transport. Many physical and biological systems are, however, better described by persistent random walks that switch between multiple states of ballistic motion. So far, traveling fronts in persistent random walk models have only been analyzed in special, simplified cases. Here, we formulate the general model of reaction-transport processes in such systems and show how to compute the expansion velocity for arbitrary number of states. For the two-state model, we obtain a closed-form expression for the velocity and report how it is affected by different transport and replication parameters. We also show that nonzero death rates result in a discontinuous transition from quiescence to propagation. We compare our results to a recent observation of a discontinuous onset of propagation in microtubule asters and comment on the universal nature of the underlying mechanism. \\
\end{abstract}


\maketitle

\section{Introduction}
Transport and self-replication give rise to propagating fronts found across many systems in physics, chemistry, engineering, and ecology\cite{Mendez:2010ui, vanSaarloos:2003vr, Cross:1993el, Murray:2002uo, Okubo:wf, Othmer:1988ee}. Examples range from opinion spreading\cite{Wang:vp} and forest fires\cite{Mendez:forestfires} to tumor growth\cite{Gatenby5:cancer} and cell biological processes\cite{Ishihara:2014gpa, Chang:2013bc}. Diffusion is a convenient and widely-used model of the transport process\cite{vanSaarloos:2003vr, Murray:2002uo, okubo:diffusion}. This approach is often criticized because it neglects correlations or persistence between the motion at different times\cite{Codling:review}. A less known, but perhaps more crucial, assumption behind the diffusion approximation is that the velocity of particles has no upper bound\cite{Zaburdaev:2015gw, Holmes:1993ue}. Indeed, the Gaussian distribution extends infinitely far from the starting position of a random walker. Such instantaneous transport could yield inaccurate or simply unphysical results in many situations of front propagation \cite{Holmes:1993ue, Mendez:2010ui}. 

The simplest way to account for a finite transport velocity is to consider a particle that moves with a fixed velocity and reorients its direction at a constant rate. Such motion is known as persistent, correlated, or ballistic random walks and is described by the generalized telegraph equation in the continuum limit\cite{Goldstein:1951,Holmes:1993ue, Codling:review, Dunbar:td, Stage:2016Levy}. Propagating fronts of self-replicating persistent random walks have been studied by a variety of methods including the Wentzel-Kramers-Brillouin approximation\cite{Fedotov:2002}, phase portrait analysis \cite{Murray:2002uo,Holmes:1993ue}, Hamilton-Jacobi approach\cite{Mendez:2010ui}, and numerical simulations\cite{Vergni:stochastic_PRW}. These studies showed that the diffusion approximation predicts front velocities that exceed the maximal velocity of the particle, while the telegraph equation correctly captures the upper bound on the front velocity \cite{Holmes:1993ue, Mendez:2010ui}.

Most of the previous work focused on particles with just two states (left and right moving) and further assumed certain symmetries for reproduction, switching, and movement rates, which are not expected to hold in general \cite{Holmes:1993ue, Mendez:2007ug, Mendez:2014dh}. Here, we develop a generalized framework for traveling fronts formed by persistent random walks with an arbitrary number of states. Switching between multiple states with different motility and replication is a realistic representation of the individual behavior of bacteria\cite{Kearns:2010, Rather:2005}, motile cells\cite{LiangCox:persistentCellMotion, Campos:2010hc}, cancer cells\cite{Gerlee:2012ku}, and animals\cite{Tkacik:animalb2018,Holmes:1993ue}.
We show how to formulate an appropriate model, identify relevant parameter combinations, and compute the velocity of a traveling front. Similar to prior studies, our results are limited to the so-called ``pulled waves'' whose kinetics is controlled by the replication at the leading edge of the front\cite{vanSaarloos:2003vr, goldenfelid:pulled, aronson:weinberger}.

For the two-state model, our method produces an explicit formula for the front velocity in terms of elementary functions. Unlike previous studies, we obtained the solution for the most general model, which contains ten distinct parameters. The exact solution greatly facilitates the exploration of the parameter space and clearly shows how different
growth and transport states contribute to the overall motion of the
front. We compare the exact solution to the diffusion approximation and pay special attention to important limiting cases. In particular, we derive conditions necessary for the onset of front propagation, determine the relative importance of different replication modes, and explain asymptotic behavior at small and large replication rates. 

In many applications, the particles can not only replicate, but also die. Because this possibility has been rarely considered in the literature\cite{Lin:bacteria_bloodflow}, we thoroughly analyze the effect of death on front propagation. Our main finding is that the onset or cessation of propagation is accompanied by a discontinuous jump of the velocity. That is, as the replication rate increases, the velocity jumps from zero to a finite value when the population transitions from net death to net growth. 

Velocity jumps have been recently observed in growing microtubule networks\cite{Ishihara:2016eh}. Similar to persistent random walks, microtubules and other biopolymers possess long-lived states in which the polymer either polymerizes or depolymerizes\cite{Mitchison:1984ub, Grimm:2003cp, Dogterom:1993tz}. Many of such biopolymers also catalyze their own nucleation and thus self-assemble as traveling fronts\cite{Ishihara:2014ih, Ishihara:2016eh, Buttenschon:2019ct}. We provide an intuitive explanation for velocity jumps in biopolymers and highlight the important differences between self-replicating particles and self-replicating polymers. 

\section{Phenomenological description: Diffusion equation with advection and growth} \label{primerDiffRxn}

To gain a conceptual understanding, we first describe a simple model of coupled growth and transport: a reaction-advection-diffusion equation,

\begin{equation}
\dfrac{\partial n}{\partial t}=D\dfrac{\partial^{2}n}{\partial x^{2}}-U\dfrac{\partial n}{\partial x}+g(n) n\textrm{,}\label{eq:rxndiffeqn}
\end{equation}

where $n$ is the population density, $D$ is the diffusion coefficient, $U$ is the advection velocity, and $g(n)$ is the per capita growth rate. The advection velocity is typically set to zero in the context of range expansions and biological invasions because there is no preferred direction for dispersal. However, in some situations, such as bacteria in the urinary tract or aquatic animals in rivers, advection velocity cannot be neglected because it accounts for the directed motion due to fluid flow in the environment\cite{Ballyk:1999, Pachepsky:upstream}. 

When~$g(0)>0$, any initial inoculation results in local growth until the density saturates at~$\hat{n}$ such that~$g(\hat{n})=0$. In addition to local growth, the population spreads spatially producing two traveling fronts on each side of the population. These fronts translate in space with constant velocity~$V$ without changing their shape. In other words,~$n(t,x)$ depends only on~$x-Vt$ rather than~$t$ and~$x$ separately. Here, we exclusively focus on pulled waves\cite{vanSaarloos:2003vr,Birzu:2018jz}, i.e expansions whose velocity is controlled only by the dynamics at the leading edge. Pulled expansions occurs when growth dynamics have negative or weakly positive feedback. For example, expansion is guaranteed to be pulled when $g(n)$~is monotonically decreasing\cite{vanSaarloos:2003vr,Birzu:2018jz}. The front velocities of pulled waves are obtained by linearizing equation~(\ref{eq:rxndiffeqn}) and are given by

\begin{equation}
V=U\pm2\sqrt{g(0)D}\textrm{,}\label{eq:rxndiff_frontvelocity}
\end{equation}

Each of the two fronts may move either leftward or rightward depending on the relative magnitude of the advection velocity~$U$ and the so-called Fisher velocity~$2\sqrt{g(0)D}$.

\begin{figure}
\begin{centering}
\includegraphics[scale=0.5]{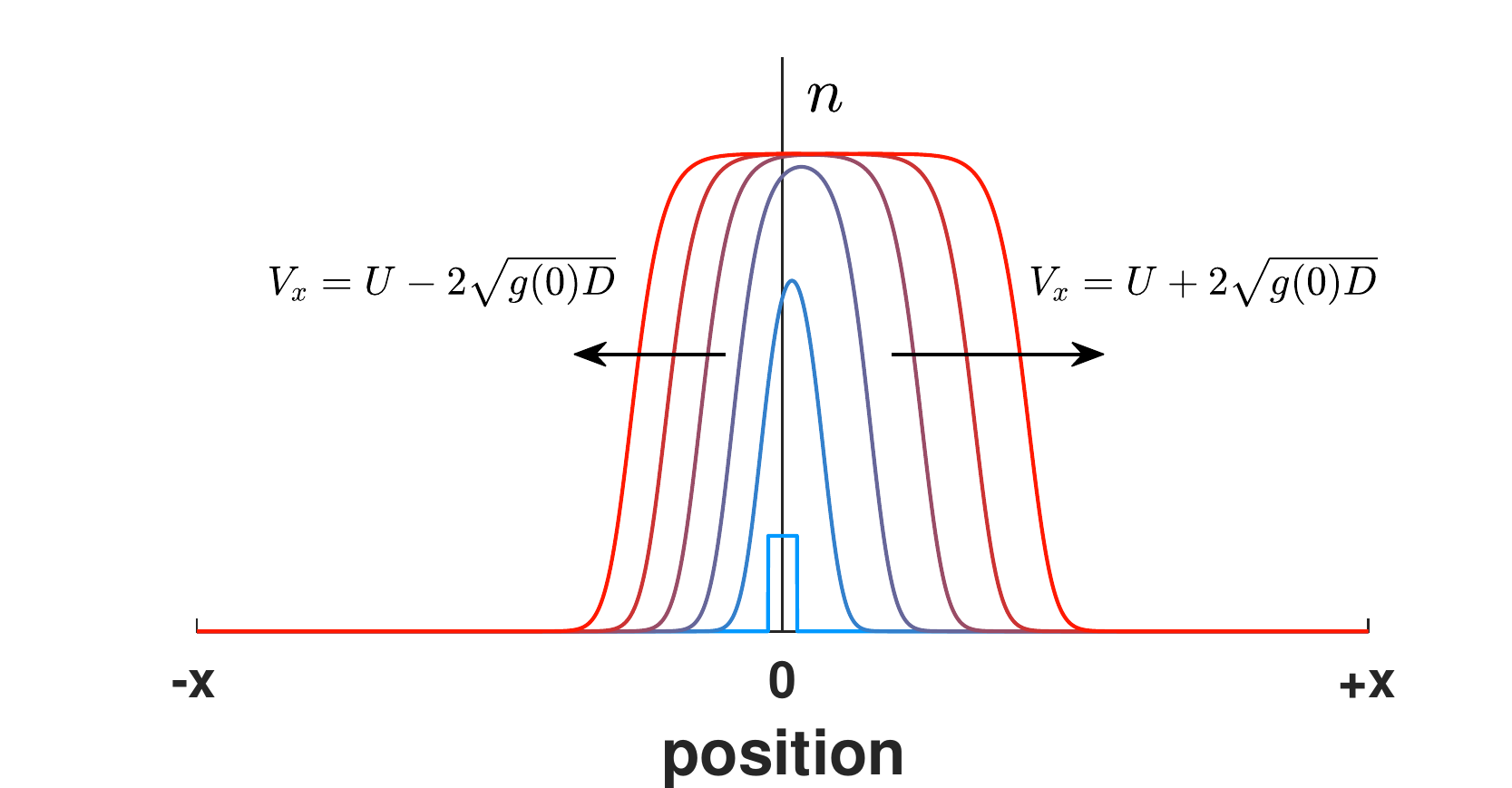}
\par\end{centering}
\caption{Solution of the diffusion-advection-reaction equation. After a transient, two expansion fronts are established on each side of the population. Time is shown with color. \label{fig:evolDiffRxnEqn}}
\end{figure}

To fix ideas, let us consider logistic growth with death, which is commonly used in mathematical biology,

\begin{equation}
g(n)=r\left(1-\frac{n}{K}\right)-d\textrm{.}\label{eq:growthtermexample}
\end{equation}

Here,~$r$ is the growth rate at low population density, $K$~is the carrying capacity, and $d$~is the death rate. From equation~(\ref{eq:rxndiff_frontvelocity}), one can easily determine how front velocities depend on the model parameters and, for example, compute the critical growth rate~$r^{\mathrm{*}}=\frac{U^2}{4D}$, which is necessary to propagate against the flow. 

The onset of propagation can be either gradual when~$V$ changes continuously as model parameters are varied or sudden when~$V$ jumps from zero to a non-zero value. The gradual onset occurs only when~$d=0$, which is the case that has received most attention in the literature. A sudden onset is the generic case because it occurs whenever the death rate is positive. Indeed, imagine increasing the growth rate~$r$ from below~$d$ to above~$d$. When~$r<d$, the population size declines until the population becomes extinct. When~$r>d$, the population grows, with the population translating with nonzero velocity~$U$ as soon as it is viable. In the present context, this observation is rather trivial, but, in more complex system such as biopolymers, it leads to important experimental signatures that can be used to elucidate biological function or mechanism~\cite{Ishihara:2016eh}. 

This manuscript is concerned with developing the theory of the aforementioned phenomena in self-replicating persistent random walks (Sec.~\ref{different_replication_modes}) and polymers (Sec.~\ref{subsec:polymerformulation}).

\section{A general framework for persistent random walk of replicating particles with multiple states}

As we discussed in the introduction, diffusion approximation permits arbitrary speeds of movement. Indeed, equation~(\ref{eq:rxndiff_frontvelocity}) predicts unbounded growth of~$V$ with~$g(0)$. This unphysical behavior can be eliminated by enforcing an upper bound on the particle velocity. Persistent random walk is the simplest and widely applicable models that constrains particle velocity. In this model, particles exist in multiple states each with a fixed velocity. The transitions between the states occur with constant, but potentially state-dependent rates.

Here, we formulate a general framework for the persistent random walk of replicating particles
with an arbitrary number of states, and provide the exact solution
for the front velocity. Throughout the paper, we
use the term ``states'' to refer to particles that exhibit different
velocity, switching, and replication rates. Depending on the specific
application, ``states'' could be considered
as species, phenotypes, particles, or behaviors.

\subsection{Mechanistic formulation}
Consider a one-dimensional system of replicating particles that can be in one of~$N$ distinct states. Under the assumptions stated above, the dynamics of this system is given by~$N$ coupled first-order differential equations:

\begin{equation}
\dfrac{\partial n_{\alpha}}{\partial t}=-v_{\alpha}\dfrac{\partial n_{\alpha}}{\partial x}-d_{\alpha}(n)\cdot n_{\alpha}+\sum_{\beta=1}^{N}f_{\alpha\beta}(n)\cdot n_{\beta}+\sum_{\beta=1}^{N}r_{\alpha\beta}(n)\cdot n_{\beta}\label{eq:biologicalformulation}
\end{equation}

where $n_{\alpha}$ is the density of particles in state~$\alpha$; $v_{\alpha}$ is the
particle velocity; $d_{\alpha}$ is the death rate;~$f_{\alpha\beta}\geq0$ is the switching rate from state~$\beta\ne\alpha$ into state~$\alpha$; and~$r_{\alpha\beta}\geq0$ is the replication rate at which particles is state~$\beta$ produce particles in state~$\alpha$. The switching rate out of state~$\alpha$ is given by~$-f_{\alpha\alpha}$, which also equals~$\sum_{\gamma\ne\alpha}f_{\gamma\alpha}$ by the conservation of probability.  

In general, all the rates can depend on all the population densities of the particles~$\{n_\alpha\}$. For pulled waves, the expansion velocities depend only on the values of these rate at low population densities~($n_\alpha\to0$), so we omit the density dependence in the following and implicitly assume that the values are taken in the limit of vanishing~$n_\alpha$. In simulations, we need to impose density-dependence to prevent unbounded growth. We accomplished by setting~$r_{\alpha\beta}(n)=r_{\alpha\beta}(1-n/K)$, where~$n=\sum_{\alpha}n_\alpha$ is the total population density and~$K$ is the carrying capacity. All other rates had no density-dependence in our simulations.

\subsection{Reduced formulation}

The last three terms in equation~(\ref{eq:biologicalformulation}) have the same functional form and can be combined for a more compact formulation

\begin{equation}
\dfrac{\partial n_{\alpha}}{\partial t}=-v_{\alpha}\dfrac{\partial n_{\alpha}}{\partial x}+\sum_{\beta=1}^{N}\Lambda_{\alpha\beta}(n)\,n_{\beta}\label{eq:compactformulation}
\end{equation}

where

\begin{equation}
\Lambda_{\alpha\beta}(n)=-d_{\alpha}(n)\delta_{\alpha\beta}+r_{\alpha\beta}(n)+f_{\alpha\beta}(n)\textrm{,}\label{eq:compactformulation2}
\end{equation}

and~$\delta_{\alpha\beta}$ denotes Kronecker's delta. Note that the choice of $\Lambda$ is somewhat constrained because the mechanistic rates:~$d_{\alpha}$,~$r_{\alpha\beta}$, and $f_{\alpha\beta}$ need to be positive.

From the reduced formulation, it is clear that front velocities and other quantities will depend on the mechanistic rates only through~$\Lambda_{\alpha\beta}$. In many situations, it is however important to understand how changes in specific mechanistic rates affect population dynamics. Thus, we need a way to decompose~$\Lambda_{\alpha\beta}$ into its constitutive parts. This decomposition is, however, not unique.
When needed, we overcome this ambiguity by imposing additional requirements. Specifically, we require that~$\underset{\beta}{\sum}f_{\alpha\beta}=0$, i.e. switching between states conserves the number of particles and set~$f_{\alpha\beta}=\Lambda_{\alpha\beta}$ for~$\alpha\ne\beta$. The diagonal terms~$g_{\alpha}$ then accounts for both death and replication, which is assumed to produce only particles of the same type as the parent. Thus,~$\Lambda_{\alpha\beta}=g_{\alpha}\delta_{\alpha\beta}+f_{\alpha\beta}$,
where $g_{\alpha}$ can take take both positive and negative values. 

\subsection{Asymptotic solution for pulled fronts}

Velocity of the traveling fronts is the key property of a population of self-replicating persistent random walks. Without loss of generality, we focus on the velocity of the right edge of the population. This velocity can be computed following the procedure described in Ref.~\cite{vanSaarloos:2003vr}. The detailed derivation is provided in Appendix \ref{subsec:Detailedsolution}, but the outline of the calculation is summarized below. As noted earlier, our solution
applies to pulled fronts whose velocity can be computed by linearizing the dynamical equation and setting $\Lambda_{\alpha\beta}$ to~$\Lambda_{\alpha\beta}(0)$.

We seek the long-time asymptotics for the traveling front, characterized by a constant velocity and shape. Following linearization, the dynamical equation is Fourier transformed in space and Laplace transformed in time. The Fourier variable is denoted as~$k$ and Laplace variable as~$s$. The resulting system of linear algebraic equations is solved by the standard methods, and the inverse Fourier and Laplace transforms are performed. The resulting integrals are evaluated in the long time limit using saddlepoint approximation. 

The key quantity that arises in the above-described calculation is the following matrix

\begin{equation}
B_{\alpha\beta}(\kappa)=\Lambda_{\alpha\beta}-v_{\alpha}\kappa\delta_{\alpha\beta},\label{eq:defineB}
\end{equation}

where parameter~$\kappa$ is real~(the relevant Fourier mode has purely imaginary~$k$, so we introduced~$\kappa=-ik>0$). The eigenvalues of this matrix, $s^{(l)}$, describe available saddle points. The saddle point that predicts the largest velocity dominates the long-time dynamics.

For each eigenvalue, the corresponding front velocity is determined by solving

\begin{equation}
\frac{ds^{(l)}}{d\kappa}=\frac{s^{(l)}(\kappa)}{\kappa}\label{eq:dsdk}
\end{equation}

with respect to~$\kappa$. We denote the solution of equation~(\ref{eq:dsdk}) by~$\kappa^{(l)}_f$. This quantity specifies the spatial decay rate of all particle densities in the comoving reference frame~$n_{\alpha}\sim e^{-\kappa^{(l)}_f (x-Vt)}$. Once,~$\kappa_f^{(l)}$ is known the front velocity is given by

\begin{equation}
V^{(l)}=\frac{s^{(l)}(\kappa_{f}^{(l)})}{\kappa_{f}^{(l)}}.\label{eq:VeqSK}
\end{equation}

The actual velocity and spatial decay rate are given by~$l$ that predicts the largest $V^{(l)}$: 

\begin{equation}
V=\underset{l}{\textrm{max}}\,V^{(l)}.\label{eq:takemaxV}
\end{equation}

For a two-state system with velocities~$v_1$ and~$v_2$, the front velocity of the right side of the population can be computed in closed form:

\begin{equation}
V=\frac{(\Lambda_{22}v_{1}-\Lambda_{11}v_{2})+\mathsf{sgn}(v_{1}-v_{2})(\Lambda_{22}v_{1}+\Lambda_{11}v_{2})\sqrt{\frac{\Lambda_{12}\Lambda_{21}}{\Lambda_{12}\Lambda_{21}-\Lambda_{11}\Lambda_{22}}}}{(\Lambda_{22}-\Lambda_{11})+\mathsf{sgn}(v_{1}-v_{2})(\Lambda_{11}+\Lambda_{22})\sqrt{\frac{\Lambda_{12}\Lambda_{21}}{\Lambda_{12}\Lambda_{21}-\Lambda_{11}\Lambda_{22}}}}
\label{eq:ExactSolutionFrontLambda}
\end{equation}

where $\mathsf{sgn} $ is the sign function. The detailed derivation of this result is provided in Appendix \ref{subsec:DetailedsolutionTwoStateLambda}. We note that the above formula contains a lot of nontrivial information. Indeed, the model has six independent parameters and only one of them could be set to unity by rescaling time.

In the absence of growth and death, our result for~$V$ reduces to the average velocity~($V=\bar{v}$), which is defined as follows
\begin{equation}
\bar{v}=\frac{v_{1}f_{2}+v_{2}f_{1}}{f_{2}+f_{1}},
\label{eq:vBar}
\end{equation}

where~$f_1$ denotes~$f_{21}$ and~$f_2$ denotes~$f_{12}$, i.e.~$f_1$ and~$f_2$ are the rates of switching out of states~$1$ and~$2$ respectively.

\section{Persistent random walk of replicating particles}
\begin{figure}
\begin{centering}
\includegraphics[scale=0.46]{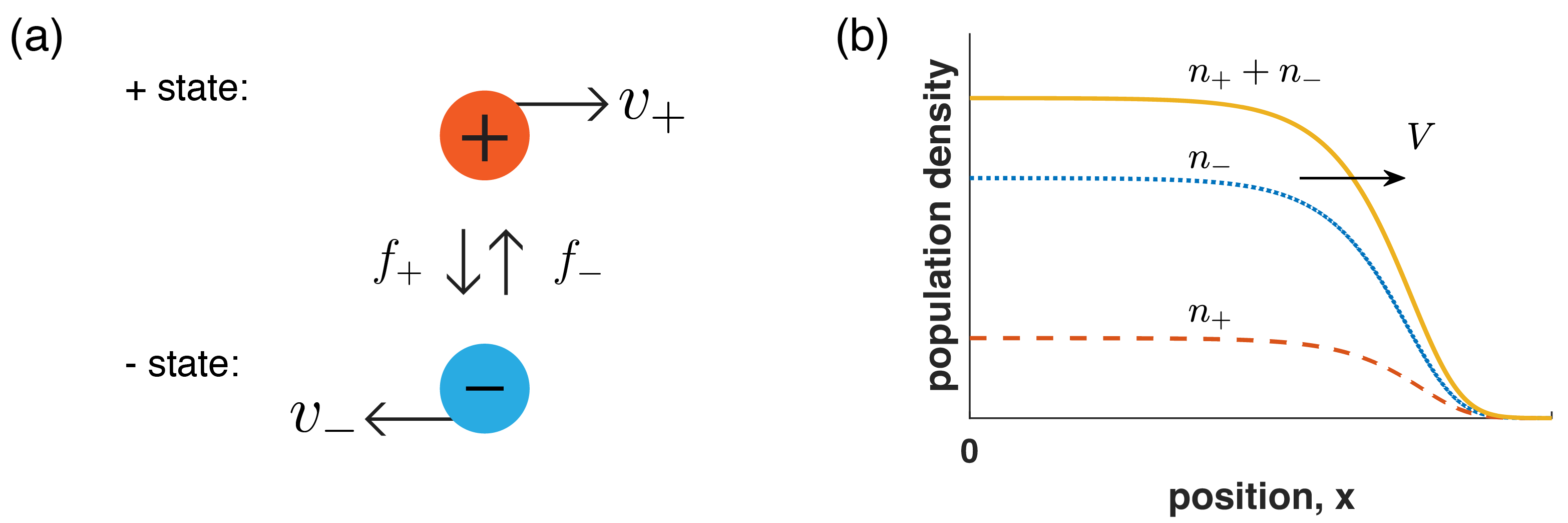}
\par\end{centering}
\caption{Two-state model of persistent random walks with replication.
(a) Illustration of the dynamics described by equation~(\ref{eq:TwostateRpp}). Particles interconvert between two
states: a ``$+$ state'' that moves to the right at speed $v_{+}$
and a ``$-$ state'' that moves to the left at speed $v_{-}.$ The
transition frequencies are denoted by $f_{+}$ and $f_{-}$.
(b) Numerical solution for the shape of the right front in the model with the following parameters
$v_{+}=20,\,v_{-}=10,\,f_{+}=3,\,f_{-}=1,\,r_{++}=1,\,r_{+-}=r_{-+}=r_{--}=0$
and $d_{+}=d_{-}=0$. \label{fig:twostate_cartoon_threeOverlay}}
\end{figure}

To understand the implications of equation~(\ref{eq:ExactSolutionFrontLambda}), it is convenient to recast the model into an intuitively interpretable form of persistent random walks; see figure~\ref{fig:twostate_cartoon_threeOverlay}a. We assume that random walks can move either rightward or leftward with speeds~$v_{+}$ and~$v_{-}$. Switching between the two states occurs at rates $f_{+}\geq0$ and $f_{-}\geq0$. In addition, particles can replicate and die with state-dependent rates. This leads to the following re-parameterization of the generic model defined above:

\begin{equation}
\begin{aligned}
v_{1} &= v_{+}\\
v_{2} &= -v_{-}
\end{aligned}
\qquad\textrm{and}\qquad
\begin{aligned}
\Lambda_{11} &= r_{++}-d_{+}-f_{+} &\\
\Lambda_{22} &= r_{--}-d_{-}-f_{-} &\\
\Lambda_{12} &= r_{+-}+f_{-} &\\
\Lambda_{21} &= r_{-+}+f_{+}. &
\end{aligned}
\label{eq:Lambda2state}
\end{equation}

The population densities are denoted as~$n_+$ and~$n_-$, and their dynamics is described by:

\begin{equation}
\left\{ \begin{alignedat}{2}\dfrac{\partial n_{+}}{\partial t} & = & -v_{+}\dfrac{\partial n_{+}}{\partial x}-f_{+}n_{+}+f_{-}n_{-}+r_{++}n_{+}+r_{+-}n_{-}-d_{+}n_{+}\\
\dfrac{\partial n_{-}}{\partial t} & = & +v_{-}\dfrac{\partial n_{-}}{\partial x}+f_{+}n_{+}-f_{-}n_{-}+r_{-+}n_{+}+r_{--}n_{-}-d_{-}n_{-}
\end{alignedat}.
\right.\label{eq:TwostateRpp}
\end{equation}

The front velocity is obtained by substituting equation~(\ref{eq:Lambda2state}) into equation~(\ref{eq:ExactSolutionFrontLambda}).

Compared to the simple reaction-diffusion-advection equation, persistent random walk model has more parameter. Instead of advection velocity and diffusion constant, there are two velocities and two switching rates. Furthermore, there are four replication rates instead of one. Below we describe how each of these parameters influences front propagation and compare the results to the prediction of the simpler reaction-diffusion-advection model equation~(\ref{eq:rxndiffeqn})~(to which we refer to as the diffusion approximation). 

Most of the discussion below is based on the exact solution. To check its validity, we performed a few numerical simulations~(see Appendix \ref{subsec:numericalsimulations} for details). Numerical results are presented in figure~\ref{fig:compareDifflimitvsExact_Vr} to \ref{fig:compareDifflimitvsExact_Vf}, which show excellent agreement between the theory and simulations. 

 Given the large number of parameters it is important to develop some intuition first. To this end, we begin by considering a simple model with a single replication rate $r_{++}>0$, and all other replication and death rates equal to zero. This simplified case is used to illustrate the effects of persistence on front propagation and compare the exact solution and the diffusion approximation. Then, we return to the most general case and determine how various elements of the replication matrix influence front propagation. Finally, we analyze the effect of death rates. 

\subsection{Comparison between the exact solution and the diffusion limit}

With only~$r_{++}$ not equal to zero, the expression for the front velocity simplifies to

\begin{equation}
V=\dfrac{2v_{-}\sqrt{f_{+}r_{++}}-(f_{+}+r_{++})v_{-}+f_{-}v_{+}}{f_{-}+f_{+}+r_{++}-2\sqrt{f_{+}r_{++}}}.\label{eq:VrpponlyMaintext}
\end{equation}

When~$r_{++}=0$, the velocity is given by the time-averaged velocity of the random walker~$\bar{v}$. This average velocity also equals to the advection velocity~$U$ in the diffusion approximation. To avoid confusion and simplify the notation, we refer to both of these velocities as~$\bar{v}$. The detailed derivation of the diffusion approximation is provided in Appendix \ref{subsec:appendixDiffLimit}.

\begin{figure}
\centering{}\includegraphics[scale=0.4]{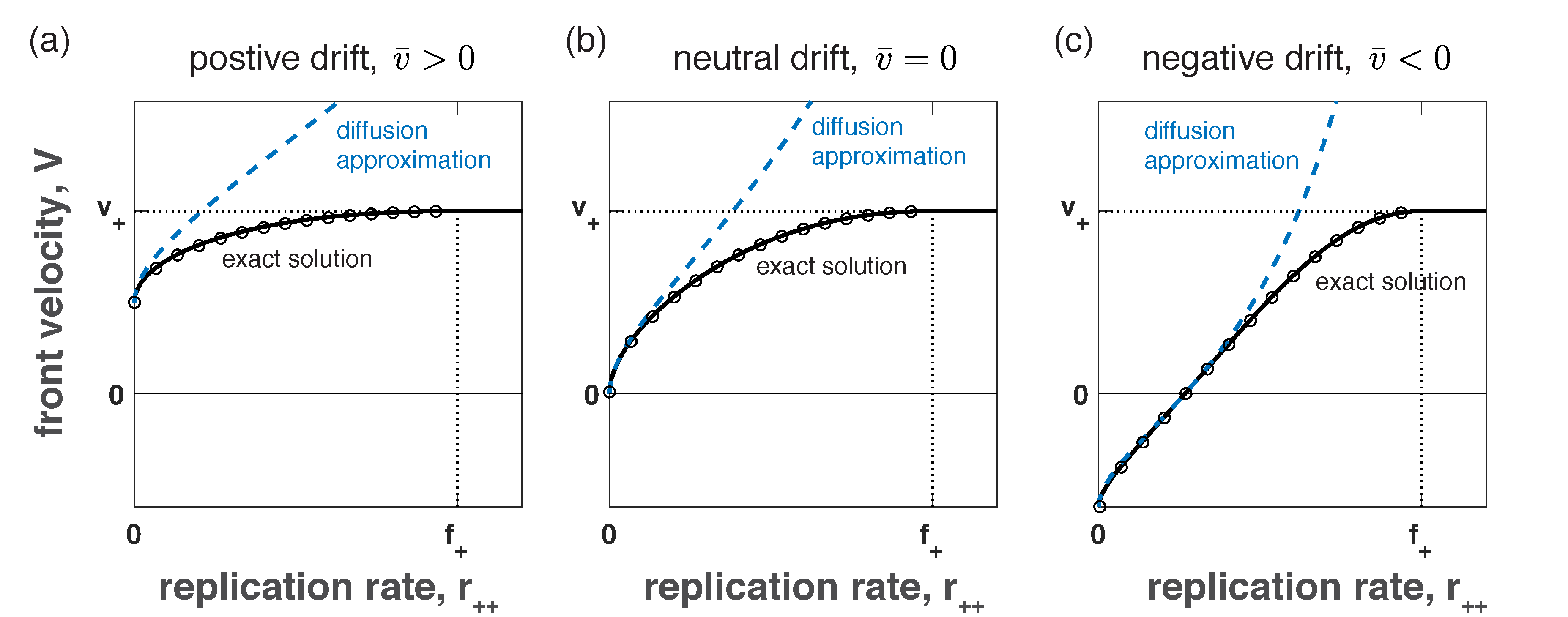}\caption{Dependence of the front velocity on the replication rate. We plot the results of numerical simulations~(circles), our exact solution (solid lines, equation~(\ref{eq:VrpponlyMaintext})), and the diffusion approximation
(dashed lines, equation~(\ref{eq:rxndiff_frontvelocity})). The effective parameters in the diffusion approximations are derived in Appendix~\ref{subsec:appendixDiffLimit}. The following parameters are the same in all three panels: $v_{+}=10,\,v_{-}=10,\,f_{+}=3,\,r_{++}>0,\,r_{+-}=r_{-+}=r_{--}=0$. The remaining parameter~$f_{-}$ is chosen such that~$\bar{v}>0$ in (a), $\bar{v}=0$ in (a), $\bar{v}<0$ in (c). Specifically, $f_{-}=9$ in~(a), $f_{-}=3$ in (b), and~$f_{-}=0.7$ in (c). Note that a minimal replication rate is required for the front to propagate rightward when~$\bar{v}<0$ and that~$V=v_{+}$ when~$r_{++}\ge f_{+}$.
\label{fig:compareDifflimitvsExact_Vr}}
\end{figure}

The major difference between the exact solution and the diffusion approximation is their dependence on $r_{++}$. In the diffusion approximation, the front velocity grows indefinitely with $r_{++}$, but the exact solution correctly recapitulates the physical constraint that $V\leq v_{+}$. We illustrate this difference in figure \ref{fig:compareDifflimitvsExact_Vr} where we show $V(r_{++})$ for three parameter combinations that correspond to $\bar{v}>0$, $\bar{v}=0$, and~$\bar{v}<0$. Diffusion approximation works best for $\bar{v}$ near zero and small replication rates. Deviations from this sweet spot lead to drastic differences, and the diffusion approximation can both overestimate (figure \ref{fig:compareDifflimitvsExact_Vr}a) and underestimate (figure \ref{fig:compareDifflimitvsExact_Vr}c) the velocity of the front.

In contrast to replication, the dependence of $V$ on $v_{+}$ and $v_{-}$ is largely determined by dimensional analysis and Galilean invariance (see Appendix \ref{appendixGalilean}). In particular, $V$ must be a linear combination of $v_{+}$ and $v_{-}$ of the following form:

\begin{equation}
V=v_{+}-w(\Lambda_{\alpha\beta})\cdot(v_{+}+v_{-}).\label{eq:GalileanInvarianceMain}
\end{equation}

\begin{figure}
\centering{}\includegraphics[scale=0.4]{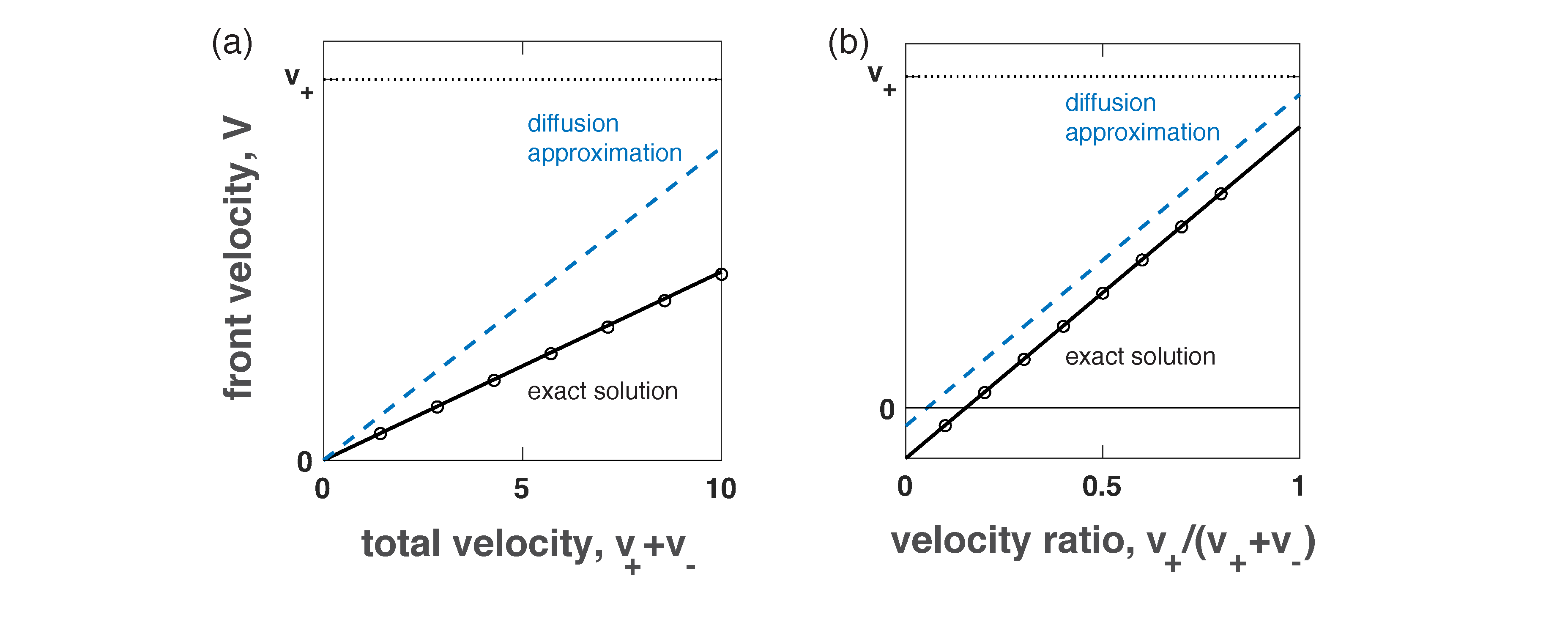}\caption{Dependence of the front velocity on the particle velocities. (a) shows the front velocity as a function
of~$v_{+}+v_{-}$. We kept $f_{+}=3$, $f_{-}=3$, $r_{++}=1$,
and $v_{+}/v_{-}=3/7$ constant, while increasing $v_{+}+v_{-}$.
(b)~shows the front velocity as a function of~$\frac{v_{+}}{v_{+}+v_{-}}$.
We kept $f_{+}=3$, $f_{-}=3$, $r_{++}=1$, and $v_{+}+v_{-}=10$
constant, while varying the relative magnitude of $v_{+}$ and $v_{-}$.
The diffusion approximation is shown with dashed lines~(equation (\ref{eq:rxndiff_frontvelocity})) and our exact solution with solid lines~(equation (\ref{eq:VrpponlyMaintext})). The simulation results are shown with circles. The effective parameters in the diffusion approximations are derived in Appendix~\ref{subsec:appendixDiffLimit}.
\label{fig:compareDifflimitvsExact_Vv}}
\end{figure}

Both the exact solution and the diffusion approximation comply with equation~(\ref{eq:GalileanInvarianceMain}) and therefore should show similar dependence on the microscopic velocities. figure \ref{fig:compareDifflimitvsExact_Vv} confirms this expectation. When the ratio of $\frac{v_{+}}{v_{-}}$ is held fixed (figure \ref{fig:compareDifflimitvsExact_Vv}a), the expansion velocity is linear in $v_{+}+v_{-}$ for both equation (\ref{eq:rxndiff_frontvelocity}) and (\ref{eq:VrpponlyMaintext}) although the slopes of the lines do differ. When the sum of the velocities is held fixed (figure~\ref{fig:compareDifflimitvsExact_Vv}b), the dependence on the velocity ratio is also linear, but now the diffusion approximation and the exact solution differ only in the intercept, again consistent with equation~(\ref{eq:GalileanInvarianceMain}).

\begin{figure}
\centering{}\includegraphics[scale=0.4]{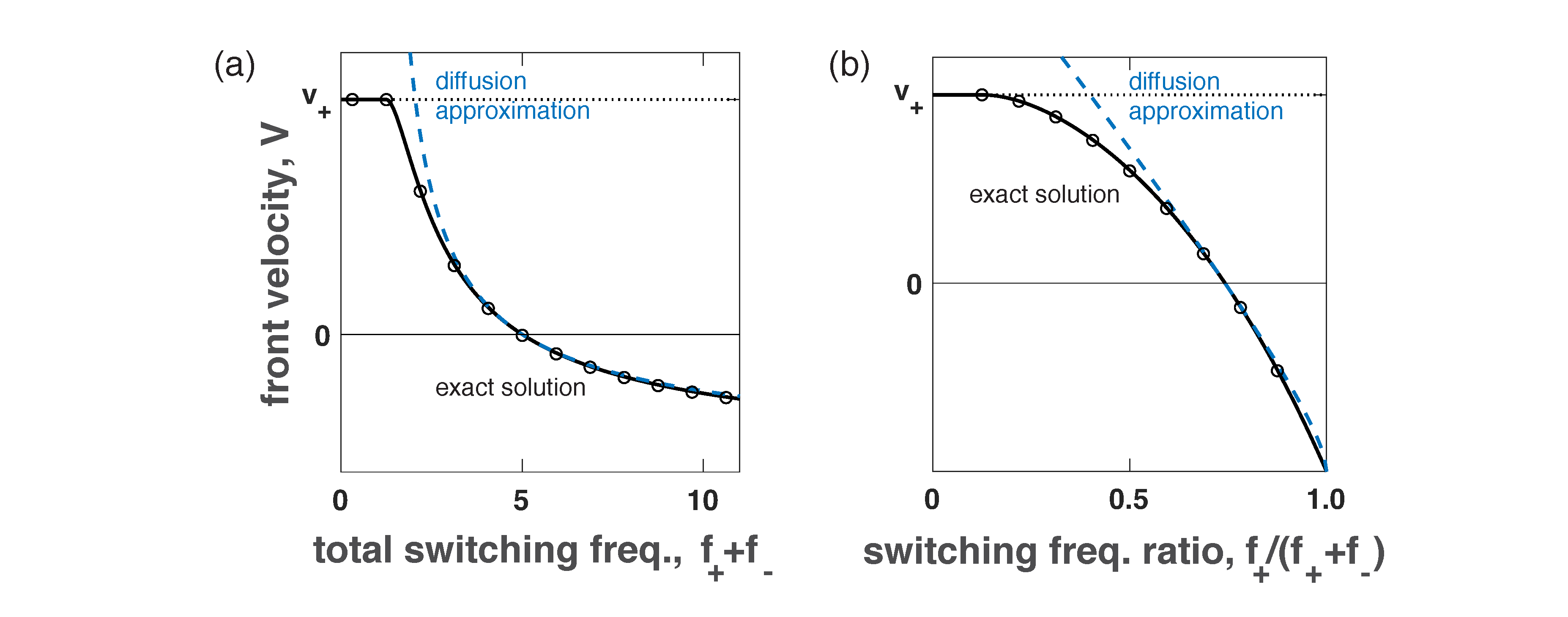}\caption{Dependence of front velocity on switching frequencies. (a) shows the front velocity as a function
of~$f_{+}+f_{-}$. We kept $v_{+}=10$, $v_{-}=10$, $r_{++}=1$,
and $f_{+}/f_{-}=4$ constant, while increasing $f_{+}+f_{-}$.
(b) shows the front velocity as a function of~$\frac{f_{-}}{f_{+}+f_{-}}$.
We kept $v_{+}=10$, $v_{-}=10$, $r_{++}=1$, and $f_{+}+f_{-}=8$ constant,
while varying $\frac{f_{-}}{f_{+}+f_{-}}$. The diffusion approximation is shown with dashed lines (equation~(\ref{eq:rxndiff_frontvelocity})) and our exact solution with solid lines (equation (\ref{eq:VrpponlyMaintext})). The simulations are shown with circles. The effective parameters in the diffusion approximations are derived in Appendix~\ref{subsec:appendixDiffLimit}.
\label{fig:compareDifflimitvsExact_Vf}}
\end{figure}

The effects of switching frequencies are somewhat similar to those of~$r_{++}$ because $V$ depends only on~$f_{+}/r_{++}$ and~$f_{-}/r_{++}$. Both the exact solution and diffusion approximation predict that the front velocity decreases with increasing the total switching frequency (figure \ref{fig:compareDifflimitvsExact_Vf}a) and the relative magnitude of $f_{+}$ (figure \ref{fig:compareDifflimitvsExact_Vf}b). The values of the velocity can, however, be very different. We find the largest deviations between the diffusion approximation and the exact solution for small switching frequencies. In this regime, the diffusion approximation predicts velocities above $v_{+}$ just as for large replication rates~(compare to figure \ref{fig:compareDifflimitvsExact_Vr}).

\begin{figure}
\begin{centering}
\includegraphics[scale=0.43]{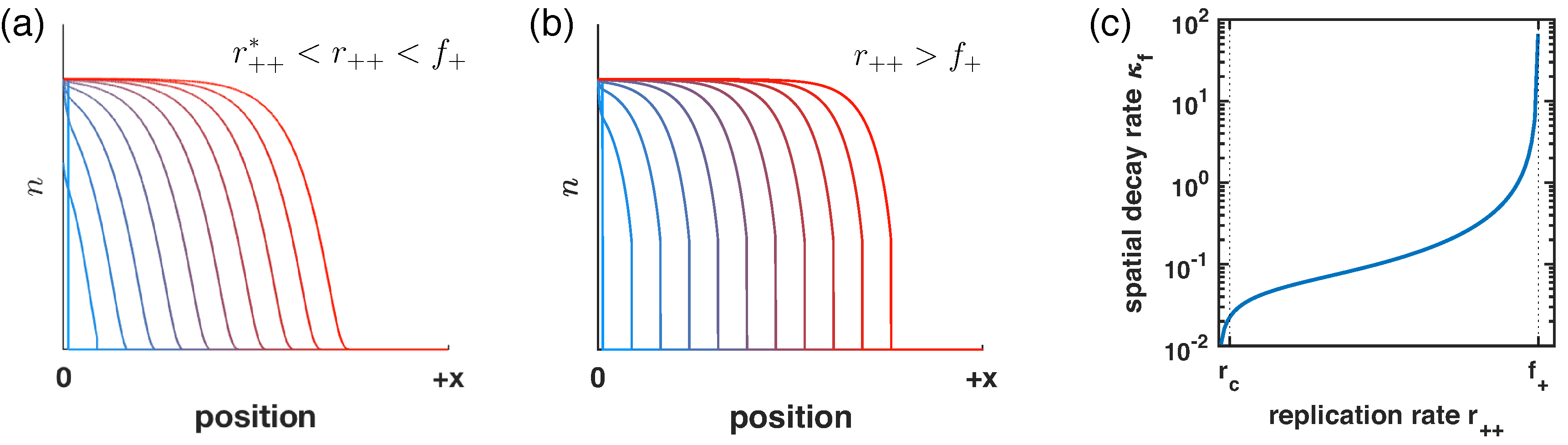}
\par\end{centering}
\caption{Shape of the expansion front. (a) shows the sigmoidal shape of the front for~$r_{++}<f_{+}$ while (b)~shows a jump in the population density at the leading edge for~$r_{++}>f_{+}$. The transition between sigmoidal and sharp front is further described in~(c), which shows the spatial decay rate~$\kappa_f$ as a function of~$r_{++}$. Here, $v_{+}=20,\,v_{-}=10,\,f_{+}=3,\,f_{-}=1$; the replication rate $r_{++}$ equals~$2.5$~in~(a) and $5$~in~(b). 
\label{fig:rEffectonFrontShape}}
\end{figure}

Finally, we show two qualitatively different front shapes in figure~\ref{fig:rEffectonFrontShape}. For~$r_{++}<f_{+}$, the front has sigmoidal shape with exponential decay at the leading edge. Qualitatively, this is the same as in the diffusion approximation. For~$r_{++}>f_{+}$, the leading edge of the front is infinitely sharp, i.e. it is effectively a step function followed by gradual saturation towards the carrying capacity. The origin of this singularity is that the front velocity saturates at~$v_{+}$ and the shape of the leading edge does not relax, but preserves any discontinuities present in the initial condition~(we start simulations with a nonzero density within an interval). 

\subsection{Contribution of different replication modes to front propagation}
\label{different_replication_modes}
In the preceding section, we considered replication only in the plus state. Now, we systematically compare how different modes of replication contribute to the onset of propagation and the velocity of the front. Our results follow directly from the exact solution; the details of the calculations are presented in Appendix \ref{subsec:velocitiesForSingleReplicationRate}. 

\begin{figure}[h]
\centering{}\includegraphics[scale=0.39]{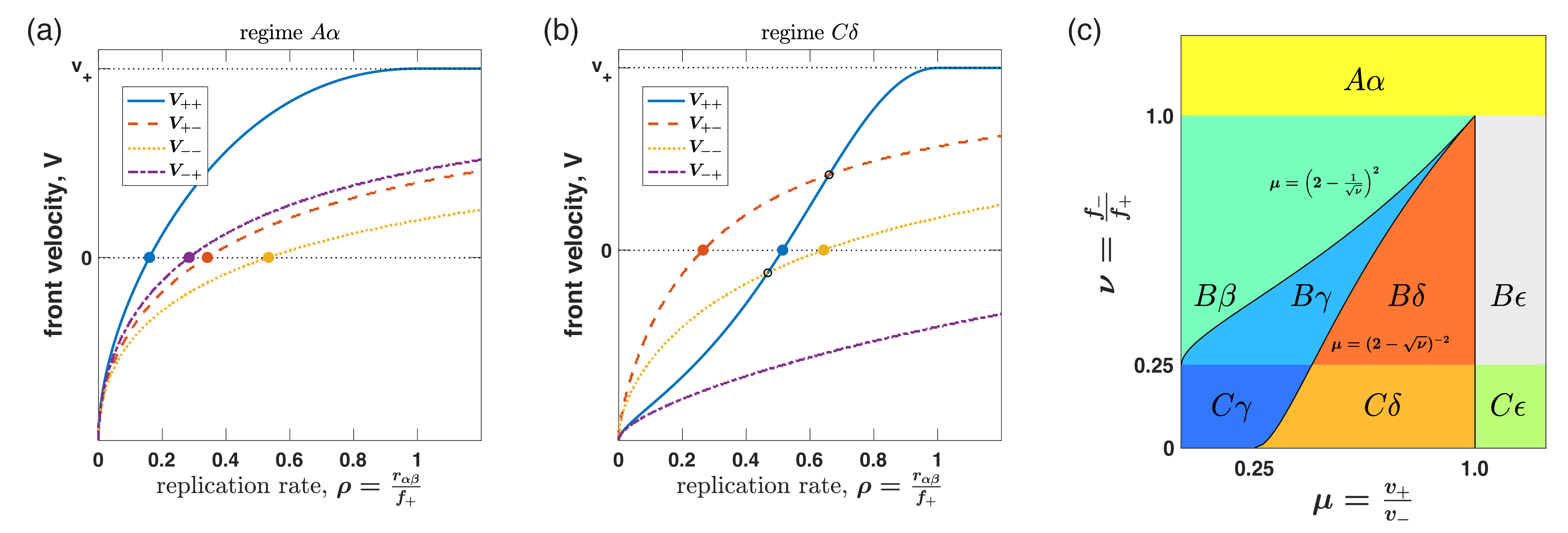}
\caption{
Ordering of front velocities and critical replication rates depends on the mode of replication.
Front velocities $V_{\alpha\beta}$ as a function of normalized replication rates $\rho=\frac{r_{\alpha\beta}}{f_{+}}$ exhibit simple~(regime $A\alpha$, panel (a)) or complex~(regime $C\delta$, panel (b)) dynamics depending on parameter values.
Filled circles indicate the critical replication rate for each mode of replication.
Open circles indicate when one velocity surpasses another.
Panels (a) and (b) were obtained by keeping $v_{-}=1$ and $f_{+}=1$, but changing $\mu=\frac{v_{+}}{v_{-}}$ and $\nu=\frac{f_{-}}{f_{+}}$ to $(\mu, \nu)=(0.3, 1.2)$ and $(\mu, \nu)=(0.8, 0.1)$, respectively.
Examples of all possible regimens are shown in figure~\ref{fig:Ordering_all8regimes} in Appendix \ref{subsec:velocitiesForSingleReplicationRate}.
Phase diagram (c) depicts parameter space for all possible regimes of velocity orderings. The explanation of the regime labeling scheme is provided in Tab.~\ref{tab:orderingTable}.}
\label{fig:diffrab}
\end{figure}

Specifically, we consider four possibilities with only one of the four replication rates~($r_{++}$, $r_{+-}$, $r_{-+}$, $r_{--}$) not equal to zero. The dependence of corresponding velocities~($V_{++}$, $V_{+-}$, $V_{-+}$, $V_{--}$) on the replication rates is shown in figure \ref{fig:diffrab}. While each~$V_{\alpha,\beta}(r_{\alpha,\beta})$ is monotonically increasing, the pace of this increase depends on the replication mode in a non-trivial way. For some parameter values, there is a clear ordering of the velocities~(figure~\ref{fig:diffrab}a, regime $A\alpha$). For other parameter values, there are transitions in the ordering of the front velocities and the critical replication rates~(figure~\ref{fig:diffrab}b, regime $C\delta$). Below, we compare the ordering of the front velocities and characterize their behavior for small and large replication rates and near the onset of propagation~($V_{\alpha\beta}$=0).

For small replication rates, we find that $V_{\alpha\beta}\approx U + C_{\alpha,\beta}\sqrt{r_{\alpha,\beta}}$; see Appendix \ref{subsec:exactSolutionsSingleRep}. The constants~$C_{\alpha,\beta}$ coincide for the $V_{++}$ and $V_{-+}$ and for $V_{+-}$ and~$V_{--}$. This clustering into two pairs is evident both in figure~\ref{fig:diffrab}a and in figure~~\ref{fig:diffrab}b.

In the opposite limit of high replication rates, the front velocity reaches the physically maximal value of $v_{+}$ (see Appendix \ref{subsec:exactSolutionsSingleRep}). While $V_{+-}$ , $V_{--}$, and $V_{-+}$ approach $v_{+}$ gradually as $r_{\alpha\beta}\rightarrow\infty$, $V_{++}$ becomes equal to~$v_{+}$ when $r_{++}=f_{+}$. In the former cases, right moving particles need to pass through left-moving state during replication, so~$V<v_{+}$ because some time is spent in the left-moving state. In the latter case, however, some particles always stay in the right-moving state when the production rate~$r_{++}$ exceeds the loss rate~$f_{+}$, so they create a front moving with~$V=v_{+}$.

When~$U<0$, front velocities have different sign for small and large replication rates. The condition~$V_{\alpha,\beta}=0$ marks the onset of propagation strictly to the right, which is important in many applications~\cite{Mendez:2007ug}. Below we provide critical replication rates~$r^{*}_{\alpha\beta}$ necessary for a right-moving front.

\begin{itemize}
\item only $r_{++}\ne0$  

\begin{equation}
r_{++}^{*}=\left(\sqrt{f_{+}}-\sqrt{\dfrac{v_{+}f_{-}}{v_{-}}}\right)^{2}\label{eq:rppcritical}
\end{equation}

\item only $r_{+-}\ne0$  

\begin{equation}
r_{+-}^{*}=\dfrac{(f_{+}v_{-}-f_{-}v_{+})^{2}}{4f_{+}v_{+}v_{-}}\label{eq:rmpcritical}
\end{equation}

\item only $r_{--}\ne0$

\begin{equation}
r_{--}^{*}=\left(\sqrt{\dfrac{v_{-}f_{+}}{v_{+}}}-\sqrt{f_{-}}\right)^{2}=\dfrac{v_{-}}{v_{+}}r_{++}^{*}\label{eq:rmmcritical}
\end{equation}

\item only $r_{-+}\ne0$ 

\begin{equation}
r_{-+}^{*}=\dfrac{(f_{+}v_{-}-f_{-}v_{+})^{2}}{4f_{-}v_{+}v_{-}}=\dfrac{f_{+}}{f_{-}}r_{+-}^{*}\label{eq:rpmcritical}
\end{equation}
\end{itemize}

In all cases, parameters that favor front propagation in the positive direction lower the critical replication rate.

Finally, we determine how the ordering of the relative magnitudes of~$V_{\alpha,\beta}$ as well as $r^{*}_{\alpha\beta}$ depend on model parameters. In Appendix~\ref{subsubsec:pairwise}, we analyse all six pairwise comparisons of the four velocity equations. We first find that the ordering of front velocities depends the dimensionless parameter $\mu=\frac{v_{+}}{v_{-}}$, which defines regimes A, B, and C with unique ordering and crossing of velocities. The ordering of critical replication rate additionally depends on the value of $\nu=\frac{f_{-}}{f_{+}}$, which leads to additional possibilities labeled by $\alpha$, $\beta$, $\gamma$, $\delta$, $\epsilon$. In total, there are eight possible regimes: $A\alpha$, $B\beta$, $B\gamma$, $B\delta$, $B\epsilon$, $C\gamma$, $C\delta$, and $C\epsilon$ (see figure~\ref{fig:Ordering_all8regimes} and Table~\ref{tab:orderingTable} in Appendix~\ref{subsubsec:pairwise}). These results are summarized visually by a phase diagram in figure~\ref{fig:diffrab}b.

\subsection{Particle death leads to a discontinuous jump in the front velocity}
\label{death_rate}
\begin{figure}
\includegraphics[width=\linewidth]{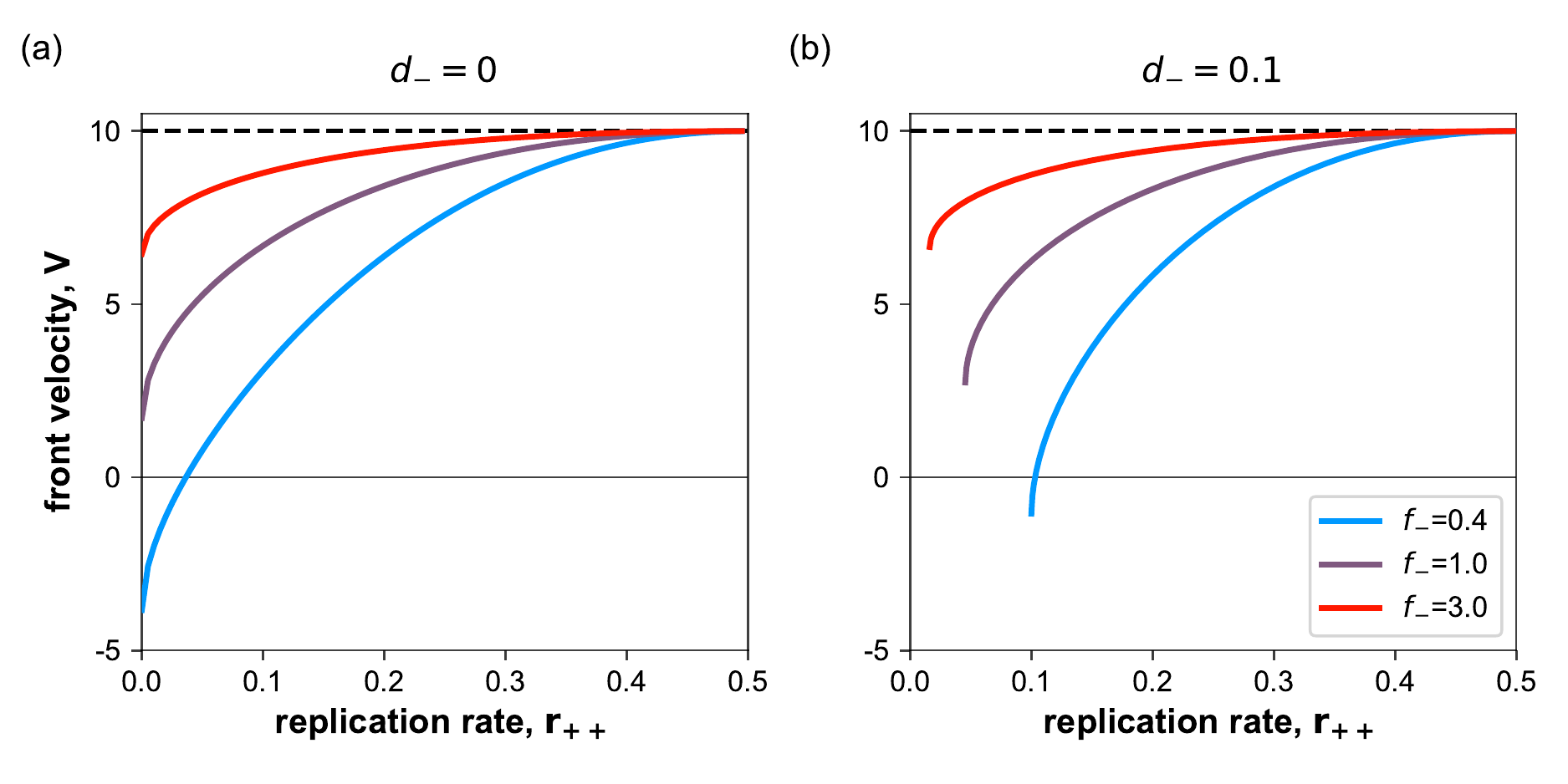}
\caption{Death causes an abrupt onset of propagation at a nonzero velocity. (a)~For a two-state system without a dying state, the front velocity exists for all positive growth rates. (b)~For one dying and one growing state, the the onset of propagation can occur abruptly with a finite minimal velocity. Here,~$v_+=10,~v_{-}=-15,~f_{+}=0.5$ . \label{fig:death_rates}}
\end{figure}

In this section, we show that nonzero death rates could lead to a qualitatively new behavior: the system can transition abruptly from extinction to a traveling wave moving at a finite velocity (figure~\ref{fig:death_rates}). This transition occurs when the net growth rate is positive in one, but negative in the other state. 

The transition between extinction and growth does not depend on particle velocities. Indeed, let us integrate the linearized equation~(\ref{eq:compactformulation}) over space: 

\begin{equation}
\int_{-\infty}^{\infty} \frac{\partial n_{\alpha}}{\partial t}  dx= \int_{-\infty}^{\infty}  -v_{\alpha}\dfrac{\partial n_{\alpha}}{\partial x}+\sum_{\beta=1}^{N}\Lambda_{\alpha\beta} \,n_{\beta} \,dx .
\end{equation}

The boundary terms disappear because the concentration is zero at $\pm \infty$, and we obtain the non-spatial version of the model:

\begin{equation}
\frac{d C_{\alpha}}{d t}= \sum_{\beta=1}^{N} \Lambda_{\alpha\beta}  \,C_{\beta}, \label{eq:well_mixed}
\end{equation}

where $C_{\alpha} (t)= \int_{-\infty}^{\infty} n_{\alpha}(t,x) dx$.

Extinction~($C_{\alpha}=0$) occurs only when all the eigenvalues of~$\Lambda_{\alpha,\beta}$ have negative real part. For the two-species system, the transition from extinction to growth occurs when one of the eigenvalues crosses zero. Thus, generically, we expect the transition when the trace is negative~$\Lambda_{22} + \Lambda_{11}<0$ and the determinant is zero:

\begin{equation}
\Lambda_{22}\Lambda_{11}- \Lambda_{21}\Lambda_{12}=0. \label{eq:rcrit_condition}
\end{equation}

We illustrate the transition to growth further using the following simple, yet generic, scenario with~$r_{++}>0$,~$d_{-}>0$, and~$r_{+-}=r_{-+}=d_{+}=r_{--}=0$. The minimal replication rate $r^{\mathrm{gap}}_{+}$ is given by
\begin{equation}
r^{\mathrm{gap}}_{+}= f_{+} - \frac{f_{+} f_{-} }{d_{-}+f_{-}}.
\label{eq:rcrit}
\end{equation}

Note that~$r^{\mathrm{gap}}_{+}$ is different from the critical replication rates equation (\ref{eq:rppcritical})-(\ref{eq:rpmcritical}) for which the front velocity is zero.
At the transition, the velocity of the front is given by

\begin{equation}
V^{\mathrm{gap}}=\frac{ \Lambda_{22}v_{+}- \Lambda_{11}v_{-} }{ \Lambda_{11}+\Lambda_{22}}= \frac{ -(d_{-}+f_{-})v_{+}-( r_{+}-f_{+} ) v_{-} }{ r^{\mathrm{gap}}_{+}-f_{+}-d_{-}-f_{-} }.
\label{eq:gapv_death}
\end{equation}

Following Ref.~\cite{Ishihara:2016eh}, we term this minimal velocity as the gap velocity since it characterizes the gap in the possible values that front velocity can take. 
The diffusion limit predicts both~$r^{\mathrm{gap}}_{+}$ and~$V^{\mathrm{gap}}$ correctly; see equations~(\ref{eq:difflimitRgap}) and~(\ref{eq:difflimitVgap})  .

\section{Origins of gap velocity in reaction-transport systems}

Onset of growth with a non-zero velocity has been recently reported in a growing network of microtubules\cite{Ishihara:2016eh}. Based on their model, the authors of Ref.~\cite{Ishihara:2016eh} argued that the velocity gap is a unique feature of persistent motion with replication. Here, we re-examine this claim and identify the key differences between populations of persistent random walks and persistently-growing polymers. We first review the polymer model and compare its behavior to diffusion approximation and replicating particles that do not die. We then revisit this analysis by including death rates to account for the possibility of complete polymer depolymerization. 

\subsection{Propagating fronts of replicating polymers undergoing persistent growth dynamics} \label{subsec:polymerformulation}

We begin by briefly summarizing the key results of the model for replicating polymers undergoing persistent growth\cite{Ishihara:2016eh}. In this model, polymers have a static that does not move and a dynamic end at which the polymer can polymerize with velocity~$v_{+}$ or depolymerize with velocity~$v_{-}$. It is assumed that the dynamic end is to the right of the static end due to the mechanical interactions between the polymers in the network. This implies that, if we define the position of the static end as~$x_s$ and the polymer length as~$l$, then the position of the dynamic end is~$x_{d}=x_s+l$. We denote the polymerizing or growing state of the dynamic end as~$+$ and the depolymerizing or shrinking sate as~$-$. The transition rate from~$+$ to~$-$ sate is~$f_{+}$. The rate of the reverse transition is~$f_{-}$. Replication in this model is the nucleation of a new zero-length polymer in the growing sate. We denote this replication rate by $Q_{(t,x)}$. 

The mathematical formulation of the above-described polymer model reads

\begin{equation}
\left\{ \ \begin{aligned}\frac{\partial \rho_{+}(t,x,l)}{\partial t}=-v_{+}\frac{\partial \rho_{+}}{\partial l}-f_{+}\rho_{+}+f_{-}\rho_{-} & +Q\cdot\delta(l),\\
\frac{\partial \rho_{-}(t,x,l)}{\partial t}=+v_{-}\frac{\partial \rho_{-}}{\partial l}+f_{+}\rho_{+}-f_{-}\rho_{s,}
\end{aligned}
\right.\label{eq:polymerPDEsystem}
\end{equation}

where $\rho_{+}(t,x,l)$ and $\rho_{-}(t,x,l)$ are the densities of polymers in growing and shrinking states respectively at time~$t$ with their static ends located at~$x$ and length equal to~$l$. We use $\delta(l)$ to denote the Dirac delta function.

The basic model of polymer replication considered in Ref.~\cite{Ishihara:2016eh} assumes that each growing end can branch. This makes~$Q$ a function of only the density of the growing ends,~$C_{+}(t,x_{+})$, which can be computed from~$\rho_{+}(x_{s},l)$ by taking an integral over~$x_{s}$ and~$l$ while keeping~$x_{s}+l=x$; see Ref.~\cite{Ishihara:2016eh}. Upon assuming simple logistic-like saturation at high~$C_{+}$, we obtain the following equation for~$Q$

\begin{equation}
Q(t,x)=v_{+}\,\rho_{+}(t,\,x,\,l=0)=R_{+}\,C_{+}(t,x)\left(1-\dfrac{{C_{+}(t,x)}}{K}\right),\label{eq:nucleationbifurcation}
\end{equation}

where~$K$ controls the saturation density, and $R_{+}$ is the replication rate.

The calculations in Ref.~\cite{Ishihara:2016eh} show that this system can either become extinct or expand as a traveling wave. Note that the wave travels in~$x$-space with the distribution of polymer lengths being constant at each spatial location in the co-moving reference frame. The velocity of the front is given by

\begin{equation}
V_{\textrm{polymer}}=\frac{v_{+}(v_{+}f_{-}-v_{-}f_{+})^{2}}{\left(\begin{aligned}v_{+}(v_{+}f_{-}-v_{-}f_{+})(f_{-}+f_{+})+(v_{+}+v_{-})(v_{+}f_{-}+v_{-}f_{+})R_{+}\\
-2(v_{+}+v_{-})\sqrt{v_{+}f_{+}f_{-}r(v_{+}f_{-}-v_{-}f_{+}+v_{-}R_{+})}
\end{aligned}
\right)},
\label{eq:asterVelocity}
\end{equation}

and the critical replication rate is

\begin{equation}
R_{+}^{\textrm{gap}}=f_{+}-\dfrac{v_{+}}{v_{-}}f_{-}.
\label{eq:rCriticalPolymerPlusendstim}
\end{equation}

At the transition from extinction to expansion, the front velocity jumps from zero to the gap velocity

\begin{equation}
V^{\textrm{gap}}_{\textrm{polymer}}=\underset{R_{+}\rightarrow R_{+}^{gap}}{\lim}V_{\textrm{polymer}}=\dfrac{v_{+}v_{-}(-v_{+}f_{-}+v_{-}f_{+})}{v_{+}^{2}f_{-}+v{}_{-}^{2}f_{+}}.\label{eq:Vgap}
\end{equation}

\subsection{Comparison between propagation onset in persistent random walks and polymers}
The theory of persistent polymers predicts a minimal spreading velocity given by equation~(\ref{eq:Vgap}). Is this a qualitatively new feature of persistent polymers or can this transition be captured by persistent random walk model or even the diffusion approximation?

\begin{figure}[h]
\centering{}\includegraphics[scale=0.54]{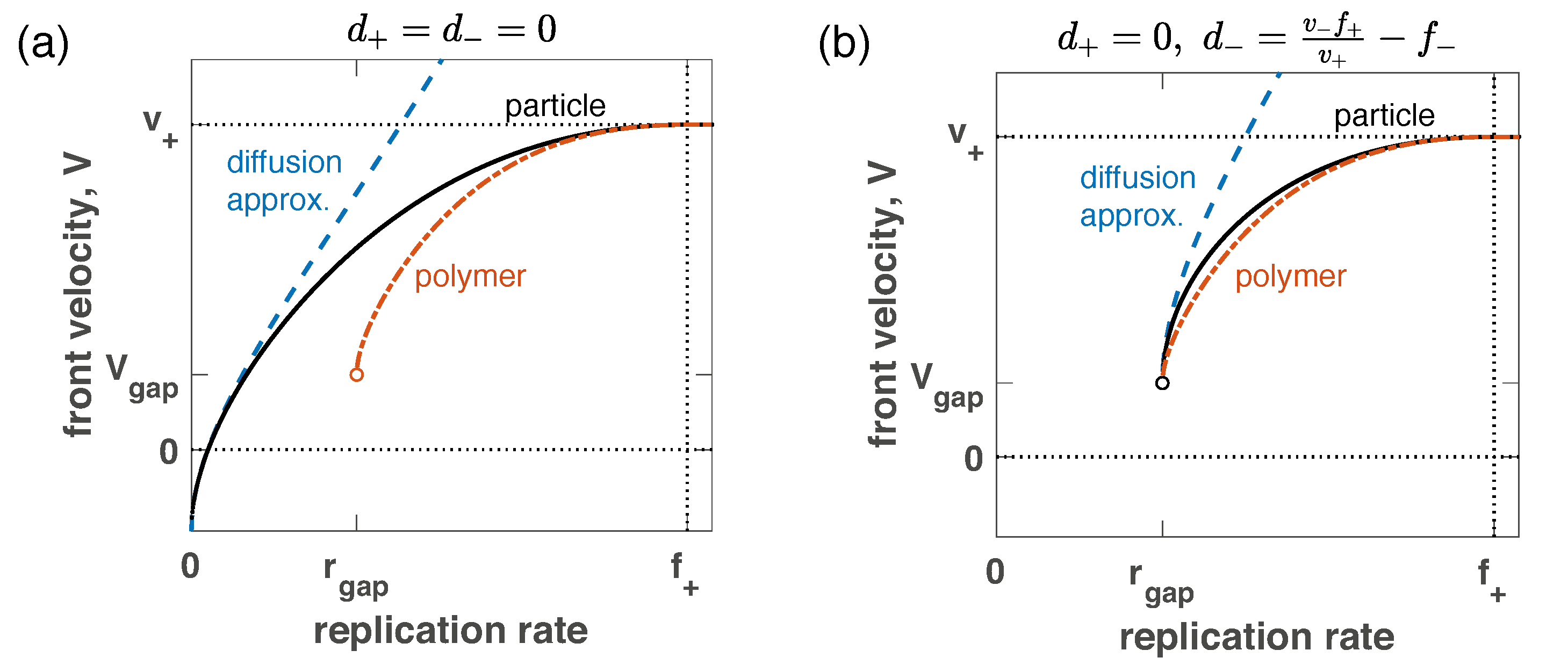}\caption{The origin of the velocity jump in polymer model. (a) shows the dependence of the front velocity on the replication rate in three models: the diffusion approximation~(dashed lines, equation (\ref{eq:difflimitV})), the persistent random walk model~(solid lines, equation (\ref{eq:VrpponlyMaintext})), and the self-replicating polymer model (dashed and dotted lines, equation (\ref{eq:asterVelocity})). Only the polymer model has a jump in front velocity when the replication rate is increased. We used the parameters of the polymer model to parameterize the other models:~$v_{+}=10$, $v_{-}=15$, $f_{+}=1$, $f_{-}=1$, $d_{-}=0.5$, $r_{+-}=r_{-+}=r_{--}=d_{+}=0$.
(b) When death is incorporated using equation~(\ref{eq:TimeShrinking}), the polymer and the random walk models show qualitatively the same behavior.
\label{fig:GapVelocityAcrossModels}}
\end{figure}

To answer this question, we first compared the predictions of the polymer model and the persistent random walk model with identical parameters. Typical dependence of the velocity on the replication rate is shown in figure~\ref{fig:GapVelocityAcrossModels}a. As suggested in Ref.~\cite{Ishihara:2016eh}, there is indeed a qualitative difference between the two models. Propagation begins with zero velocity in the random walk model, but with a nonzero gap velocity in the polymer model.

Given that we observed gap velocity for the random walk model with death, we hypothesized that the discontinuous jump in the polymer model is related to death-like events. Indeed, when polymer shrinks to zero length, it disappears from the population, i.e. it effectively dies. The effective death rate can be estimated as the inverse lifetime of the polymer, which has been computed in Ref.~\cite{Ishihara:2016eh}. Using their results, we then investigated the persistent random walk model with~$d_{+}=0$~(growing polymers never die) and  

\begin{equation}
d_{-}=f_{+}\frac{v_{-}}{v_{+}} - f_{-}.
\label{eq:TimeShrinking}
\end{equation}

The results of this comparison are shown in figure~\ref{fig:GapVelocityAcrossModels}b. Now, both the polymer and the random walk model have indistinguishable values of the critical replication rate and the gap velocity. This agreement holds for all parameter values. Indeed, one can confirm this by substituting the death rate from equation~(\ref{eq:TimeShrinking}) into equation~(\ref{eq:gapv_death}) for the gap velocity in the random walk model. The result is identical to equation~(\ref{eq:Vgap}) for the gap velocity in the polymer model. Thus, the sudden onset of growth in the polymer model can be traced to the disappearance of polymers that shrink to zero length. The persistent nature of polymer growth is less important because the diffusion approximation predicts identical values of the critical nucleation rate and the gap velocity.

 figure~\ref{fig:GapVelocityAcrossModels}b also shows that the agreement between the random walk and polymer models is only qualitative. Both reproduce the discontinuous onset of growth and saturation of the front velocity at~$v_{+}$, but the intermediate values of the front velocity do differ. This difference is however very small for many biologically relevant parameter values. Therefore, one might be able to use a much simpler and substantially more computationally efficient random walk model to approximate the dynamics of polymer networks.

\section{Discussion}

Front propagation in replicating and persistently moving particles is a widespread phenomenon that has many applications\cite{Holmes:1993ue, Mendez:2010ui}. Such reaction-transport systems can be approximately mapped onto well-understood reaction-diffusion systems, but the diffusion approximation can be rather inaccurate or even produce unphysical results. Furthermore, the diffusion approximation is insensitive to certain parameters of the persistent random walks that do affect the behavior of the system. 

In this paper, we formulated a general model of persistent random walks with an arbitrary number of motility and replication states, and provided a simple method to compute the expansion velocity. The presented framework is very flexible and can account for complex transitions between states. For example, a time delay could be modeled by requiring that any transition occurs through one or more intermediate states. 

The general framework was then applied to a two-state system. We obtained an exact solution for the most general model and explored the dependence of the front velocity on all model parameters. We found that the diffusion approximation is accurate only for a limited range of parameter values. In particular, the diffusion approximation fails for high replication or low switching rates~(figure~\ref{fig:compareDifflimitvsExact_Vr} and~\ref{fig:compareDifflimitvsExact_Vf}). 

Furthermore, the diffusion approximation does not fully capture the dependence of the front velocity on replication rates in different states. We analyzed the effect of four possible replication mechanisms: two choices for the state in which the particle replicates and two choices of the offspring state. These replication modes make unequal contributions to the front velocity, and the efficacy of each replication mode depends on other parameters. We completely characterized these dependencies and identified eight regions in the parameter space that exhibit different dependence of front velocities on replication rates; see figure~\ref{fig:diffrab}.

We also analyzed the effects of particle death, which have been rarely considered in the literature. The death rates result in somewhat unusual behavior: There is a minimal speed at which the front can propagate. This minimal speed occurs right when the replication can counterbalance death, and the population starts to grow. Such transitions from no growth to rapidly moving expansion fronts have been observed in microtubule networks~\cite{Ishihara:2016eh} and we sought to provide an intuitive explanation for their origin. We found that one can identify an effective death term in the persistent polymer model, which fully accounts for the critical replication rate and velocity jump. In fact, the persistent particle model approximates the persistent polymer model rather well across the whole range of replication rates~(figure~\ref{fig:GapVelocityAcrossModels}b). 

Overall, our work extends earlier results on persistent random walks\cite{Mendez:2010ui, vanSaarloos:2003vr, Cross:1993el, Murray:2002uo, Okubo:wf, Othmer:1988ee} to a much more general class of models and provides new insights into the role of different replication modes and death rates. These results might find applications across a number of fields where one encounters front propagation. 

\begin{acknowledgments}
KI acknowledges the support from the ELBE fellowship from the Center for Systems Biology Dresden (Max Planck Society). Overall, this work was supported by Simons Foundation Grant \#409704 (to KSK); by the Research Corporation for Science Advancement through Cottrell Scholar Award \#24010 (to KSK); by the Scialog grant \#26119 (to KSK); and by the Gordon and Betty Moore Foundation grant \#6790.08 (to KSK).
\end{acknowledgments}

\appendix

\section{Appendix}

\subsection{Calculation of the front velocity}\label{subsec:Detailedsolution}

\subsubsection{Fourier and Laplace transforms of the linearized equation}

Our goal is to compute the velocity of traveling fronts described by equation~(\ref{eq:compactformulation}), which we repeat below

\begin{equation}
\dfrac{\partial n_{\alpha}}{\partial t}=-v_{\alpha}\dfrac{\partial n_{\alpha}}{\partial x}+\sum_{\beta=1}^{N}\Lambda_{\alpha\beta}n_{\beta}\quad\textrm{with\quad\ensuremath{\Lambda_{\alpha\beta}}=\ensuremath{g_{\beta}\delta_{\alpha\beta}}+\ensuremath{f_{\alpha\beta}}}.\label{eq:compactalphabeta}
\end{equation}

First, we perform the Fourier transform in the space domain, $x\rightarrow k$:

\begin{equation}
\dfrac{\partial n_{\alpha}(t,k)}{\partial t}=-iv_{\alpha}kn_{\alpha}(t,k)+\sum_{\beta}\Lambda_{\alpha\beta}n_{\beta}(t,k).\label{eq:Fouriertransf}
\end{equation}

Then, we perform the Laplace transform in the time domain $t\rightarrow s$:

\begin{equation}
sn_{\alpha}(s,k)-n_{\alpha}(t=0,k)=-iv_{\alpha}kn_{\alpha}(s,k)+\sum_{\beta}\Lambda_{\alpha\beta}n_{\beta}(s,k).\label{eq:Laplacetransf}
\end{equation}

The last equation can be recast in the following form

\begin{equation}
\sum_{\beta}\left[(s+iv_{\alpha}k)\delta_{\alpha\beta}-\Lambda_{\alpha\beta}\right]n_{\beta}(s,k)=n_{\alpha}(t=0,k),\label{eq:inclsum1}
\end{equation}

which immediately suggests an equivalent matrix formulation:

\begin{equation}
\boldsymbol{A\,}n=n(t=0),\label{eq:matrixform}
\end{equation}

where $A_{\alpha\beta}=(s+iv_{\alpha}k)\delta_{\alpha\beta}-\Lambda_{\alpha\beta}$.
The solution of the matrix equation reads

\begin{equation}
n_{\alpha}(s,k)=\boldsymbol{A}^{-1}(s,k)\,n(t=0,k).\label{eq:invmatrixform}
\end{equation}

We now perform the inverse Laplace transform, $s\rightarrow t$:

\begin{equation}
n_{\alpha}(t,k)=\dfrac{1}{2\pi i}\intop_{-i\infty}^{i\infty}e^{st}n_{\alpha}(s,k)ds=\sum_{l}\underset{s^{(l)}}{\textrm{Res}}\left[e^{st}n_{\alpha}(s,k)\right].\label{eq:invLaplace}
\end{equation}

The residues arise when~$\textrm{det}\boldsymbol{A}=0$. The
degeneracy of $\boldsymbol{A}$ implies that there exist $h\neq0$
such that $\boldsymbol{A}h=0$. Therefore, $sh_{\alpha}=\sum(-iv_{\alpha}k\delta_{\alpha\beta}+\Lambda_{\alpha\beta})h_{\beta}$,
i.e. $s$ is an eigenvalue of $B_{\alpha\beta}=\Lambda_{\alpha\beta}-iv_{\alpha}k\delta_{\alpha\beta}$. Generically, $\boldsymbol{B}$~has~$N$ eigenvalues~$s^{(l)}$, which define distinct branches of the dispersion relations with corresponding velocities~$V^{(l)}$. We expect that the branch with the largest front velocity describes the long-time behavior of our system. 

Next, we perform the inverse Fourier transform, $k\rightarrow x$:

\begin{equation}
\begin{aligned}n_{\alpha}(t,x) & =\intop_{-i\infty}^{i\infty}\dfrac{dk}{2\pi}e^{ikx}\sum_{l=1}^{N}e^{s^{(l)}(k)t}\,\underset{s^{(l)}}{\textrm{Res}}\left[n_{\alpha}(s,k)\right]\\
 & =\sum_{l=1}^{N}\,\intop_{-i\infty}^{i\infty}\dfrac{dk}{2\pi}e^{ik(\xi+Vt)}e^{s^{(l)}(k)t}\,\underset{s^{(l)}}{\textrm{Res}}\left[n_{\alpha}(s,k)\right]\\
 & =\sum_{l=1}^{N}\,\intop_{-i\infty}^{i\infty}\dfrac{dk}{2\pi}e^{ik\xi}e^{t(s^{(l)}(k)t+ikV)}\,\underset{s^{(l)}}{\textrm{Res}}\left[n_{\alpha}(s,k)\right].
\end{aligned}
\label{eq:invFourier}
\end{equation}

In the above derivation, we assumed the generic situation of distinct eigenvalues of~$\boldsymbol{B}$, in which case all poles are first order. In the middle line, we have transformed the spatial variable
to the reference frame comoving with the front: $\xi=x-Vt$, where $V$ is the
front velocity we wish to determine.

\subsubsection{Asymptotic evaluation of the inverse Fourier transform}
The integrals in equation~(\ref{eq:invFourier}) can be evaluated in the long time limit using the steepest descent or saddlepoint method; see Ref.~\cite{vanSaarloos:2003vr}. The controlling factor is the exponential term, so we are looking for the saddlepoint of~$s(k)+ikV$. We also impose time invariance of the front by requiring that the real part of $s(k)t+ikV$ equals zero; otherwise, the particle density in the comoving frame either diverges or vanishes. Thus, we obtain

\begin{equation}
\begin{cases}
\dfrac{ds^{(l)}(k^{(l)})}{dk^{(l)}}|_{k^{(l)}=k_{f}^{(l)}}+iV^{(l)}=0,\\
\textrm{Re}(s^{(l)}(k_{f}^{(l)})+ik_{f}^{(l)}V^{(l)})=0,
\end{cases}\label{eq:saddlepointstationary}
\end{equation}

where~$k_{f}$ is the value of~$k$ at the saddle point. It is easy to see that~$k_{f}$ is purely imaginary, so we let~$k_{f}=i\kappa_{f}$. Then, the system of equations that we need to solve takes the following form

\begin{equation}
\begin{cases}
\left.\dfrac{ds^{(l)}(\kappa_{f}^{(l)})}{d\kappa}\right|_{\kappa^{(l)}=\kappa_{f}^{(l)}}=V^{(l)},\\
s(\kappa_{f}^{(l)})=\kappa_{f}^{(l)}V^{(l)},
\end{cases}\label{eq:saddlepointstationarykappa}
\end{equation}

where all variables are real. The solution of these equations defines the front velocity for each branch~$l$. The actual velocity is obtained by making the largest of~$l$ values. 

In the following, we will first solve for~$\kappa_{f}$ by eliminating~$V$ from the system of equations, which yields the following equation for~$\kappa_f$:

\begin{equation}
\frac{ ds^{(l)}}{d\kappa} = \frac{s}{k}.
\label{eq:kappa_eq}
\end{equation}

\subsection{Front velocity for the two-state persistent random walk}\label{subsec:DetailedsolutionTwoStateLambda}

We start from the compact formulation of the problem and quickly repeat the analysis outlined in the preceding section: 

\begin{equation}
\left\{ \begin{alignedat}{2}\frac{\partial n_{1}}{\partial t} & = & -v_{1}\frac{\partial n_{1}}{\partial x} & +\Lambda_{11}n_{1}+\Lambda_{12}n_{2}\\
\frac{\partial n_{2}}{\partial t} & = & -v_{2}\frac{\partial n_{2}}{\partial x} & +\Lambda_{21}n_{1}+\Lambda_{22}n_{2}
\end{alignedat}.
\right.\label{eq:nonredundantmodel}
\end{equation}

Fourier transform $x\rightarrow k$ for the spatial coordinate and the Laplace transform $t\rightarrow s$ for the temporal coordinate yields:

\begin{equation}
\left\{ \begin{alignedat}{2}sn_{1}-n_{1}(t=0,k) & = & -ikv_{1}n_{1} & +\Lambda_{11}n_{1}+\Lambda_{12}n_{2}\\
sn_{2}-n_{2}(t=0,k) & = & -ikv_{2}n_{2} & +\Lambda_{21}n_{1}+\Lambda_{22}n_{2}.
\end{alignedat}
\right.\label{nondredundantLaplace}
\end{equation}

We recast this result in the matrix form

\begin{equation}
\boldsymbol{A}\left(\begin{array}{c}
n_{1}\\
n_{2}
\end{array}\right)=\left(\begin{array}{c}
n_{1}(t=0,\kappa)\\
n_{2}(t=0,\kappa)
\end{array}\right),\label{eq:nonredundantLaplaceMatrixForm}
\end{equation}

where

\begin{equation}
\boldsymbol{A}\equiv\left[\begin{array}{cc}
s+v_{1}\kappa-\Lambda_{11} & -\Lambda_{12}\\
-\Lambda_{21} & s+v_{2}\kappa-\Lambda_{22}
\end{array}\right].\label{defineAT}
\end{equation}

We obtain~$s(k)$ by setting the determinant of~$A$ to zero

\begin{equation}
\textrm{det}(\boldsymbol{A})=(s+v_{1}\kappa-\Lambda_{11})(s+v_{2}\kappa-\Lambda_{22})-\Lambda_{21}\Lambda_{12}=0.\label{eq:detAT0}
\end{equation}

Instead of directly solving the above equation and substituting it into equation~(\ref{eq:kappa_eq}), we differentiate equation~(\ref{eq:detAT0}) first, then solve for~$ds/d\kappa$, and finally substitute the result into equation~(\ref{eq:kappa_eq}). This procedure yields additional equation

\begin{equation}
2s^{2}+2(v_{1}+v_{2})\kappa s+(-\Lambda_{11}-\Lambda_{22})s+2v_{1}v_{2}\kappa^{2}+(-v_{1}\Lambda_{22}-v_{2}\Lambda_{11})\kappa=0.\label{eq:dkdetAT}
\end{equation}

We now need to solve equation~(\ref{eq:dkdetAT}) and equation~(\ref{eq:detAT0}) simultaneously. This is can be done by constructing a linear combination of these equations to eliminate~$s^2$ and then solving a linear equation for~$s$ in terms of~$\kappa$. The result can be substituted in either equation and solved to obtain~$\kappa$. The results read

\begin{equation}
s_{1}=\frac{-(\Lambda_{22}v_{1}-\Lambda_{11}v_{2})+\mathsf{sgn}(v_{1}-v_{2})\mathsf{sgn}(\Lambda_{11}+\Lambda_{22})(\Lambda_{22}v_{1}+\Lambda_{11}v_{2})\sqrt{\frac{\Lambda_{12}\Lambda_{21}}{\Lambda_{12}\Lambda_{21}-\Lambda_{11}\Lambda_{22}}}}{(v_{1}-v_{2})\frac{\Lambda_{11}\Lambda_{22}}{\Lambda_{12}\Lambda_{21}-\Lambda_{11}\Lambda_{22}}}
\end{equation}
\begin{equation}
s_{2}=\frac{-(\Lambda_{22}v_{1}-\Lambda_{11}v_{2})-\mathsf{sgn}(v_{1}-v_{2})\mathsf{sgn}(\Lambda_{11}+\Lambda_{22})(\Lambda_{22}v_{1}+\Lambda_{11}v_{2})\sqrt{\frac{\Lambda_{12}\Lambda_{21}}{\Lambda_{12}\Lambda_{21}-\Lambda_{11}\Lambda_{22}}}}{(v_{1}-v_{2})\cdot\frac{\Lambda_{11}\Lambda_{22}}{\Lambda_{12}\Lambda_{21}-\Lambda_{11}\Lambda_{22}}}
\end{equation}

and

\begin{equation}
\kappa_{1}=\dfrac{-(\Lambda_{11}-\Lambda_{22})-\mathsf{sgn}(v_{1}-v_{2})|\Lambda_{11}+\Lambda_{22}|\sqrt{\frac{\Lambda_{12}\Lambda_{21}}{\Lambda_{12}\Lambda_{21}-\Lambda_{11}\Lambda_{22}}}}{(v_{1}-v_{2})\cdot\frac{\Lambda_{11}\Lambda_{22}}{\Lambda_{12}\Lambda_{21}-\Lambda_{11}\Lambda_{22}}}
\end{equation}
\begin{equation}
\kappa_{2}=\dfrac{-(\Lambda_{11}-\Lambda_{22})+\mathsf{sgn}(v_{1}-v_{2})|\Lambda_{11}+\Lambda_{22}|\sqrt{\frac{\Lambda_{12}\Lambda_{21}}{\Lambda_{12}\Lambda_{21}-\Lambda_{11}\Lambda_{22}}}}{(v_{1}-v_{2})\cdot\frac{\Lambda_{11}\Lambda_{22}}{\Lambda_{12}\Lambda_{21}-\Lambda_{11}\Lambda_{22}}}
\end{equation}

where $\mathsf{sgn}(v_1-v_2) $ refers to the sign of $v_1-v_2$.

The front velocity is then obtained from equation~(\ref{eq:saddlepointstationarykappa}). By examining the cases for $\Lambda_{11}+\Lambda_{22}<0$ and $\Lambda_{11}+\Lambda_{22}>0$ separately, we find that, for either case, the front velocity that corresponds to the right side of the population is

\begin{equation}
V=\frac{(\Lambda_{22}v_{1}-\Lambda_{11}v_{2})+\mathsf{sgn}(v_{1}-v_{2})(\Lambda_{22}v_{1}+\Lambda_{11}v_{2})\sqrt{\frac{\Lambda_{12}\Lambda_{21}}{\Lambda_{12}\Lambda_{21}-\Lambda_{11}\Lambda_{22}}}}{(\Lambda_{22}-\Lambda_{11})+\mathsf{sgn}(v_{1}-v_{2})(\Lambda_{11}+\Lambda_{22})\sqrt{\frac{\Lambda_{12}\Lambda_{21}}{\Lambda_{12}\Lambda_{21}-\Lambda_{11}\Lambda_{22}}}}.
\end{equation}

\subsection{Galilean invariance}\label{appendixGalilean}
The persistent random walk model must be invariant under Galilean transformation. We can use this invariance to determine how the front velocity depends on model parameters. Indeed, consider shifting the reference frame from the lab frame to the frame moving with velocity~$u$. In other words, $v_{i}\rightarrow v_{i}-u$ and $V\rightarrow V-u$. Then, we expect that

\begin{equation}
V(v_{i},\Lambda_{\alpha\beta})=u+V(v_{i}-u,\Lambda_{\alpha\beta})\label{eq:GalileanInvariance1}
\end{equation}

Differentiating this with respect to $u$, we find

\begin{equation}
\sum_{i}\dfrac{\partial V}{\partial v_{i}}=1.\label{eq:GalileanInvariance2}
\end{equation}

Solving this equation for the two-state system gives 

\begin{equation}
V=v_{+}-w(\Lambda_{\alpha\beta})\cdot(v_{+}+v_{-}),\label{eq:GalileanInvariance3}
\end{equation}

where $w$ is a function that depends only on the switching, replication, and death rates. Thus, Galilean invariance completely determines how the front velocity depends on the velocities of the states. Furthermore, equation~(\ref{eq:GalileanInvariance3}) is also very useful for simulations because one can, for example, move into a reference frame in which the velocity in the two states are equal in magnitude, but opposite in orientation. Such a choice substantially improves the accuracy of fine difference methods; for further details see Appendix~\ref{subsec:numericalsimulations} on numerical simulations.

\subsection{Diffusion approximation}\label{subsec:appendixDiffLimit}

Here, we show that the classical reaction-advection-diffusion equation (\ref{eq:rxndiffeqn}) and its front velocity (\ref{eq:rxndiff_frontvelocity}) are a limiting case of the persistent random walk model. We consider a two-state system with right and left moving states as in equation (\ref{eq:TwostateRpp}). The advection velocity, or the time-weighted average velocity in the lab reference frame, is given by

\begin{equation}
\bar{v}=\frac{v_{+}f_{-}-v_{-}f_{+}}{f_{-}+f_{+}}.
\label{eq:Jadvection}
\end{equation}

To simplify the derivation, let us change into a reference frame that moves with velocity $u=(v_{+}-v_{-})/2$. Then, the velocities of left and right moving states have equal magnitude~$v=(v_{+}+v_{-})/2$, and equation~(\ref{eq:TwostateRpp}) takes the following form

\begin{equation}
\left\{ \begin{alignedat}{2}\dfrac{\partial n_{+}}{\partial t} & = & -v\dfrac{\partial n_{+}}{\partial x}-f_{+}n_{+}+f_{-}n_{-}+r_{++}n_{+}+r_{+-}n_{-}-d_{+}n_{+}\\
\dfrac{\partial n_{-}}{\partial t} & = & +v\dfrac{\partial n_{-}}{\partial x}+f_{+}n_{+}-f_{-}n_{-}+r_{-+}n_{+}+r_{--}n_{-}-d_{-}n_{-}
\end{alignedat},
\right.\label{eq:leastdegRefframe}
\end{equation}

which via the relations (\ref{eq:Lambda2state}) is equivalent to

\begin{equation}
\left\{ \begin{alignedat}{2}\frac{\partial n_{+}}{\partial t} & = & -v\frac{\partial n_{+}}{\partial x} & +\Lambda_{11}n_{+}+\Lambda_{12}n_{-}\\
\frac{\partial n_{-}}{\partial t} & = & +v\frac{\partial n_{-}}{\partial x} & +\Lambda_{21}n_{+}+\Lambda_{22}n_{-}
\end{alignedat}.
\right.\label{eq:nonredundantmodelGalilean}
\end{equation}

From the first equation in (\ref{eq:nonredundantmodelGalilean}), we obtain

\begin{equation}
n_{-}=\dfrac{1}{\Lambda_{12}}\left[\dfrac{\partial n_{+}}{\partial t}+v\dfrac{\partial n_{+}}{\partial x}-\Lambda_{11}n_{+}\right].\label{eq:nminus}
\end{equation}
Upon substituting this in the second equation in (\ref{eq:nonredundantmodelGalilean})
and rearranging the terms, we find that
\begin{equation}
\dfrac{1}{\omega}\dfrac{\partial^{2}n_{+}}{\partial t^{2}}+\dfrac{\partial n_+}{\partial t}=\dfrac{v^{2}}{\omega}\dfrac{\partial^{2}n_{+}}{\partial x^{2}}-\dfrac{v(\Lambda_{11}-\Lambda_{22})}{\omega}\dfrac{\partial n_{+}}{\partial x}+\dfrac{\Lambda_{21}\Lambda_{12}-\Lambda_{11}\Lambda_{22}}{\omega}n_{+},\label{eq:combined4difflimit}
\end{equation}

where $\omega=-\Lambda_{11}-\Lambda_{22}$. We now assume slow temporal
evolution $t\sim x^{2}$ and neglect $\frac{\partial^{2}}{\partial t^{2}}$
term, then equation (\ref{eq:combined4difflimit}) reduces to equation (\ref{eq:rxndiffeqn})
with

\begin{equation}
D=\dfrac{v^{2}}{\omega}\approx\dfrac{(v_{+}+v_{-})^{2}}{4(f_{+}+f_{-})},\label{eq:difflimitD}
\end{equation}

\begin{equation}
U_{u}=\dfrac{v(\Lambda_{11}-\Lambda_{22})}{\omega}\approx\dfrac{v_{+}+v_{-}}{2}\cdot\dfrac{f_{-}-f_{+}}{f_{-}+f_{+}},\label{eq:difflimitJu}
\end{equation}

where $U_{u}$ is $U$ in the reference frame moving with velocity
$u$, and

\begin{equation}
\begin{aligned}g=\dfrac{\Lambda_{21}\Lambda_{12}-\Lambda_{11}\Lambda_{22}}{\omega} & \approx\dfrac{(r_{++}+r_{-+})f_{-}+(r_{+-}+r_{--})f_{+}}{f_{+}+f_{-}}-\frac{d_{+}f_{-}+d_{-}f_{+}}{f_{+}+f_{-}}.\end{aligned}
\label{eq:difflimitR}
\end{equation}

Note that the two terms in the approximate expression for $g$ are the time averaged replication and death rates.

We obtain the final result by returning to the lab reference frame, which gives

\begin{equation}
U=U_u+u=\frac{ \Lambda_{22}v_{+}- \Lambda_{11}v_{-} }{ \Lambda_{11}+\Lambda_{22}},\label{eq:difflimitJ}
\end{equation}

and the front velocity of the right side of the population

\begin{equation}
V=U+2\sqrt{gD}\approx\dfrac{v_{+}f_{-}-v_{-}f_{+}}{f_{+}+f_{-}}+2\dfrac{v_{+}+v_{-}}{f_{+}+f_{-}}\sqrt{f_{-}(r_{++}+r_{-+}-d_{+})+f_{+}(r_{+-}+r_{--}-d_{-})} .\label{eq:difflimitV}
\end{equation}


The critical replication rate corresponding to the transition from extinction to expansion occurs when $g$, defined by equation~(\ref{eq:difflimitR}), crosses zero. This is given by the condition

\begin{equation}
\Lambda_{21}\Lambda_{12}-\Lambda_{11}\Lambda_{22}=0,\label{eq:difflimitRgap}
\end{equation}
which is identical to equation~(\ref{eq:rcrit_condition}) for persistent random walks.

The corresponding gap velocity is given by equation~(\ref{eq:difflimitV}) at $g=0$. Hence, we have
\begin{equation}
V^{\mathrm{gap}}=U,\label{eq:difflimitVgap}
\end{equation}
which matches the corresponding expression for persistent random walks.

\subsection{Front propagation with a single mode of replication} \label{subsec:velocitiesForSingleReplicationRate}

\subsubsection{Front velocity solutions $V_{\alpha\beta}$}\label{subsec:exactSolutionsSingleRep}
Here, we provide the front velocity solutions when only one replication rate $r_{\alpha\beta}$ is nonzero. We denote $V(r_{++}\geq0, r_{+-}=r_{-+}=r_{--}=0)$ as~$V_{++}$ and similarly for other replication modes.
\begin{itemize}
\item $r_{++}$ only

\begin{equation}
V_{++}=\dfrac{2v_{-}\sqrt{f_{+}r_{++}}-(f_{+}+r_{++})v_{-}+f_{-}v_{+}}{f_{-}+f_{+}+r_{++}-2\sqrt{f_{+}r_{++}}}.\label{eq:Vrpponly}
\end{equation}
This is valid for $0<r_{++}<f_{+}$, while $V_{++}=v_{+}$ for $r_{++}>f_{+}$.

The behavior for small $r_{++}$ is as follows
\begin{equation}
V_{++}=\bar{v}+\dfrac{2(v_{+}+v_{-})f_{-}\sqrt{f_{+}}}{(f_{-}+f_{+})^2}\sqrt{r_{++}}+O(r_{++}), \label{eq:VrppSmallr}
\end{equation}
where $\bar{v}=\dfrac{v_{+}f_{-}-v_{-}f_{+}}{f_{-}+f_{+}}$ is the velocity at zero growth and death rates as given by equation~(\ref{eq:vBar}).

For $r_{++} \geq f_{+}$, the profile becomes infinitely sharp, and the velocity equals $v_{+}$. As $r_{++}$ approaches $f_{+}$ from below, the velocity approaches $v_{+}$ quadratically:
\begin{equation}
V_{++}= v_{+}-\frac{v_{+}+v_{-}}{4f_{-}f_{+}} \left( f_{+}-r_{++} \right)^2.
\end{equation}

\item $r_{+-}$ only

\begin{align}
\label{eq:Vrpmonly}
\begin{split}
V_{+-}=\dfrac{f_{+} (r_{+-} - \sqrt{r_{+-} (f_{-} + r_{+-})}) v_{-} + f_{-} (r_{+-} + \sqrt{r_{+-} (f_{-} + r_{+-})}) v_{+}}{-f_{+} (r_{+-} - \sqrt{r_{+-} (f_{-} + r_{+-})}) + f_{-} (r_{+-} + \sqrt{r_{+-} (f_{-} + r_{+-})})}.
\end{split}
\end{align}

For small $r_{+-}$,

\begin{equation}
V_{+-}=\bar{v}+\dfrac{2(v_{+}+v_{-})f_{+}\sqrt{f_{-}}}{(f_{-}+f_{+})^2}\sqrt{r_{+-}}+O(r_{+-})\label{eq:VrpmSmallr},
\end{equation}

For large $r_{+-} $, the velocity approaches $v_{+}$ as
\begin{equation}
V_{+-}= v_{+}-\frac{f_{+}(v_{+}+v_{-}) }{4r_{+-}}.
\end{equation}

\item $r_{--}$ only

\begin{equation}
V_{--}=\dfrac{2v_{+}\sqrt{f_{-}r_{--}}+(f_{-}+r_{--})v_{+}-f_{+}v_{-}}{f_{-}+f_{+}+r_{--}+2\sqrt{f_{-}r_{--}}}.\label{eq:Vrmmonly}
\end{equation}

For small $r_{--}$,

\begin{equation}
V_{--}=\bar{v}+\dfrac{2(v_{+}+v_{-})f_{+}\sqrt{f_{-}}}{(f_{-}+f_{+})^2}\sqrt{r_{+-}}+O(r_{+-}).\label{eq:VrmmSmallr}
\end{equation}

As $r_{--}\rightarrow\infty$, $V_{--}$ approaches $v_{+}$ as
\begin{equation}
V_{--}= v_{+}-\frac{f_{+} (v_{-}+v_{+})}{r_{--}}.
\end{equation}

\item $r_{-+}$ only 

\begin{align}
\label{eq:Vrmponly}
\begin{split}
V_{-+}=\dfrac{f_{+} (r_{-+} - \sqrt{r_{-+} (f_{+} + r_{-+})}) v_{-} + f_{-} (r_{-+} + \sqrt{r_{-+} (f_{+} + r_{-+})}) v_{+}}{-f_{+} (r_{-+} - \sqrt{r_{-+} (f_{+} + r_{-+})}) + f_{-} (r_{-+} + \sqrt{r_{-+} (f_{+} + r_{-+})})}.
\end{split}
\end{align}

For small $r_{-+}$,

\begin{equation}
V_{-+}=\bar{v}+\dfrac{2(v_{+}+v_{-})f_{-}\sqrt{f_{+}}}{(f_{-}+f_{+})^2}\sqrt{r_{-+}}+O(r_{-+}).\label{eq:VrmpSmallr}
\end{equation}

For $r_{-+} \to \infty $, the velocity approaches $v_{+}$ as
\begin{equation}
V_{-+}= v_{+}-\frac{f^2_{+}(v_{+}+v_{-}) }{4f_{-}r_{-+}}.
\end{equation}
\end{itemize}

\subsubsection{Ordering of the front velocities $V_{\alpha\beta}$}\label{subsubsec:pairwise}

Here we examine the relative magnitudes of four velocities $V_{\alpha\beta}$ arising from a single mode of replication. From the front velocity equations (\ref{eq:Vrpponly}), (\ref{eq:Vrpmonly}), (\ref{eq:Vrmmonly}), (\ref{eq:Vrmponly}), we perform a pairwise analysis and ask the following:
\begin{itemize}
\item Is one velocity $V_{\alpha\beta}$ always greater than the other? Or, as the replication rate
is increased, does one velocity surpass the other with increasing $r_{\alpha\beta}$?
\item Given that one velocity surpasses the other, what is the condition for one critical replication $r^{*}_{\alpha\beta}$ to be larger than that of the other?
\end{itemize}
The answer to the first question depends on the dimensionless parameter $\nu=\frac{f_{-}}{f_{+}}$, while that to the second question additionally depends on $\mu=\frac{v_{+}}{v_{-}}$.
We denote the non-dimensionalized replication rates as $\rho=\frac{r_{\alpha\beta}}{f_{+}}$ to directly compare different modes of replication.

\begin{enumerate}
\item $V_{++}$ vs. $V_{+-}$ \par
If $\nu>1$, $V_{++}>V_{+-}$ for all $\rho=\frac{r_{++}}{f_{+}}=\frac{r_{+-}}{f_{+}}$.\\
If $\nu<1$, $V_{++}$ surpasses $V_{+-}$ with increasing replication
\begin{equation}
\begin{cases}
V_{++}>V_{+-} & \textrm{for }\rho>\dfrac{4(-2+\sqrt{\nu}+\nu)^{2}}{(-4+\nu)^{2}},\\
V_{++}<V_{+-} & \textrm{for }\rho<\dfrac{4(-2+\sqrt{\nu}+\nu)^{2}}{(-4+\nu)^{2}},
\end{cases}\label{eq:crossVrppVrpm}
\end{equation}

with an additional condition for the ordering of critical replication rates
\begin{equation}
\begin{cases}
r^{*}_{++}>r^{*}_{+-} & \textrm{for }\mu>(2-\sqrt{\nu})^{-2},\\
r^{*}_{++}<r^{*}_{+-} & \textrm{for }\mu<(2-\sqrt{\nu})^{-2}.
\end{cases}\label{eq:rppVSrpm}
\end{equation}

\item $V_{++}$ vs. $V_{--}$

If $\nu>1$, $V_{++}>V_{--}$ for all $\rho=\frac{r_{++}}{f_{+}}=\frac{r_{--}}{f_{+}}$.\\
If $\nu<1$, $V_{++}$ surpasses $V_{--}$ with increasing replication
\begin{equation}
\begin{cases}
V_{++}>V_{--} & \textrm{for }\rho>(1-\sqrt{\nu})^{2},\\
V_{++}<V_{--} & \textrm{for }\rho<(1-\sqrt{\nu})^{2},
\end{cases}\label{eq:crossVrppVrmm}
\end{equation}

with an additional condition for the ordering of critical replication rates
\begin{equation}
\begin{cases}
r^{*}_{++}>r^{*}_{--} & \textrm{for }\mu>1,\\
r^{*}_{++}<r^{*}_{--} & \textrm{for }\mu<1.
\end{cases}\label{eq:rppVSrmm}
\end{equation}

\item $V_{++}$ vs. $V_{-+}$

$V_{++}>V_{-+}$ for all $\rho=\frac{r_{++}}{f_{+}}=\frac{r_{-+}}{f_{+}}$, $\nu$, and $\mu$.

\item $V_{+-}$ vs. $V_{--}$

$V_{+-}>V_{--}$ for all $\rho=\frac{r_{+-}}{f_{+}}=\frac{r_{--}}{f_{+}}$, $\nu$, and $\mu$.

\item $V_{+-}$ vs. $V_{-+}$

One velocity is always greater than the other
\begin{equation}
\begin{cases}
V_{+-}<V_{-+} & \nu>1,\\
V_{+-}=V_{-+} & \nu=1,\\
V_{+-}>V_{-+} & \nu<1.
\end{cases}\label{eq:crossVrpmVrmp}
\end{equation}

\item $V_{--}$ vs. $V_{-+}$ \par

If $\nu>1$, $V_{--}<V_{-+}$ for all $\rho=\frac{r_{--}}{f_{+}}=\frac{r_{-+}}{f_{+}}$ and $\mu$.\\
If $\nu<\frac{1}{4}$, $V_{--}>V_{-+}$ for all $\rho=\frac{r_{--}}{f_{+}}=\frac{r_{-+}}{f_{+}}$ and $\mu$.\\
If $\frac{1}{4}<\nu<1$,  $V_{-+}$ surpasses $V_{--}$ with increasing replication
\begin{equation}
\begin{cases}
V_{-+}>V_{--} & \textrm{for }\rho>\dfrac{4\nu(1+2(1-2\nu)\sqrt{\nu}-3\nu+4\nu^2)}{(1-4\nu)^2},\\
V_{-+}<V_{--} & \textrm{for }\rho<\dfrac{4\nu(1+2(1-2\nu)\sqrt{\nu}-3\nu+4\nu^2)}{(1-4\nu)^2},
\end{cases}\label{eq:crossVrmmVrmp}
\end{equation}

with an additional condition for the ordering of critical replication rates
\begin{equation}
\begin{cases}
r^{*}_{--}>r^{*}_{-+} & \textrm{for }\mu>\left(2-\frac{1}{\sqrt{\nu}}\right)^{2},\\
r^{*}_{--}<r^{*}_{-+} & \textrm{for }\mu<\left(2-\frac{1}{\sqrt{\nu}}\right)^{2}.
\end{cases}\label{eq:rmmVSrmp}
\end{equation}

\end{enumerate}

\begin{figure}[h]
\centering{}\includegraphics[scale=0.45]{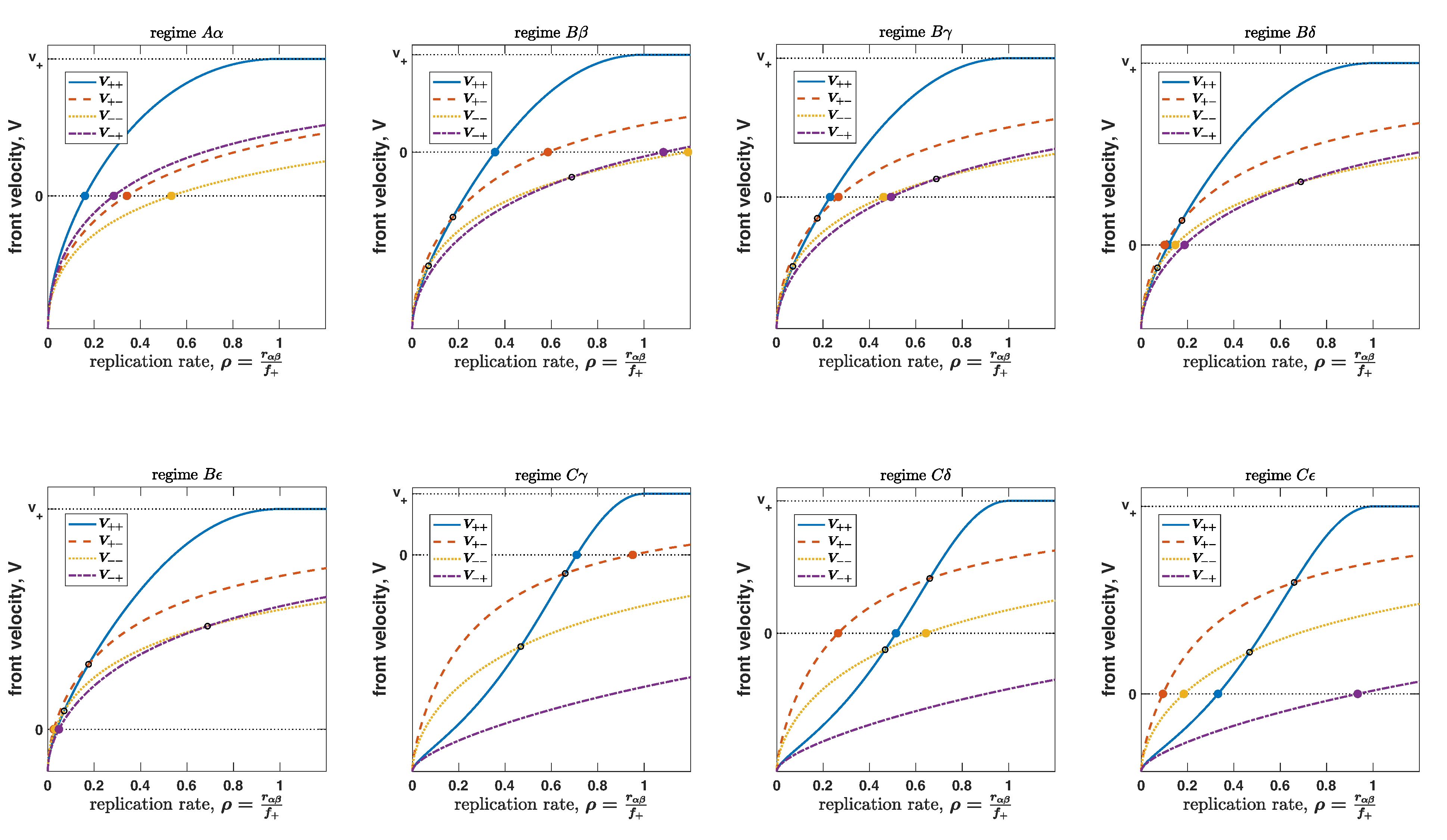}\caption{
Example of all eight regimes for the ordering front velocities $V_{\alpha\beta}$ and critical replication rates $r_{\alpha\beta}^{*}$.
Filled circles indicate the critical replication rate for each mode of replication.
Open circles indicate when one velocity surpasses another.
These velocity plots were made by keeping $v_{+}=1$ and $f_{+}=1$, while varying $\mu=\frac{v_{+}}{v_{-}}$  and $\nu=\frac{f_{-}}{f_{+}}$. Parameter values used $(\mu, \nu)$:
$A\alpha~(0.3, 1.2)$,
$B\beta~(0.3, 0.54)$,
$B\gamma~(0.5, 0.54)$,
$B\delta~(0.8, 0.54)$,
$B\epsilon~(1.2, 0.54)$,
$C\gamma~(0.25, 0.1)$,
$C\delta~(0.8, 0.1)$,
$C\epsilon~(1.8, 0.1)$.
Note that some of the critical replication rates are outside the axis limits.
\label{fig:Ordering_all8regimes}}
\end{figure}

Taken together, the six pairwise comparisons define the eight different regimes for the ordering of front velocities and critical replication rates: $A\alpha$, $B\beta$, $B\gamma$, $B\delta$, $B\epsilon$, $C\gamma$, $C\delta$, and $C\epsilon$. We provide an example of each regime in figure~\ref{fig:Ordering_all8regimes} and summarize their distinguishing features in Table~\ref{tab:orderingTable}.

\begin{table}
\begin{centering}
\begin{tabular}{|c|c|c|}
\hline 
 & \hspace{1cm} ordering of front velocities \hspace{1cm}  \tabularnewline
\hline
\hline 
$A$ & $V_{++}>V_{+-}>V_{--}>V_{-+}$ \tabularnewline
\hline
$B$ & 
$V_{++}$ surpasses $V_{+-}$,
$V_{++}$ surpasses $V_{--}$,
$V_{++}>V_{-+}$,
$V_{+-}>V_{--}$,
$V_{+-}>V_{-+}$,
$V_{--}$ surpasses $V_{-+}$ \tabularnewline
\hline
$C$ &
$V_{++}$ surpasses $V_{+-}$,
$V_{++}$ surpasses $V_{--}$,
$V_{++}>V_{-+}$,
$V_{+-}>V_{--}>V_{-+}$ \tabularnewline
\hline
\end{tabular}
\hfill \break
\hfill \break
\begin{tabular}{|c|c|c|}
\hline 
 & \hspace{1cm} ordering of critical replication rate\hspace{1cm}  \tabularnewline
\hline
\hline 
$\alpha$ & $r^{*}_{++}<r^{*}_{--}<r^{*}_{-+}<r^{*}_{+-}$ \tabularnewline
\hline
$\beta$ & $r^{*}_{++}<r^{*}_{+-}<r^{*}_{-+}<r^{*}_{--}$ \tabularnewline
\hline
$\gamma$ & $r^{*}_{++}<r^{*}_{+-}<r^{*}_{--}<r^{*}_{-+}$ \tabularnewline
\hline
$\delta$ & $r^{*}_{+-}<r^{*}_{++}<r^{*}_{--}<r^{*}_{-+}$ \tabularnewline
\hline
$\epsilon$ & $r^{*}_{--}<r^{*}_{++}<r^{*}_{+-}<r^{*}_{-+}$ \tabularnewline
\hline 
\end{tabular}
\par\end{centering}
\caption{The eight regimes are defined by ordering of front velocities and critical replication rates. The Roman labels specify the ordering of front velocities, while the Greek labels specify the ordering of critical replication rates.}
\label{tab:orderingTable}
\end{table}

\subsection{Numerical simulations}\label{subsec:numericalsimulations}
We simulated the continuum equations using an explicit first order finite difference method. To avoid numerical diffusion, the exact solution of the advection equation was used for the numerical updates. Time and space discretizations were chosen to ensure $v_{\alpha} \frac{\Delta t}{\Delta x}$ was a rational number, and concentrations of each type were periodically shifted by the appropriate integer number of lattice points. Simulations were initialized as a step function in the middle of the of the lattice. We imposed absorbing boundary concentrations at the boundaries, and ensured that simulations ended before they reached either boundary. The velocity was obtained from fitting the front position, defined as a point with concentration closest to $10^{-3}$, to a linear function of time.

\subsection{Code availability}
All relevant code (Mathematica notebooks, Python scripts, and Matlab scripts) are available online: https://github.com/keisuke-ishihara/PersistentReactionWalk.

\bibliographystyle{iopart-num}
\bibliography{mybib}

\end{document}